\documentclass[aps,prx,twocolumn,showpacs,superscriptaddress,longbibliography]{revtex4-1}
	\usepackage[bookmarks=false,linkcolor=blue,urlcolor=blue,colorlinks,citecolor=blue]{hyperref}
	\usepackage{graphicx}
	\usepackage{color}
	\usepackage{amsmath}
	 \usepackage{setspace} 
	\usepackage[normalem]{ulem}
	\usepackage[toc,page]{appendix}
\usepackage[compatibility=false,font=small,labelfont=bf, justification=justified,format=plain]{caption}
\usepackage{subcaption}
\usepackage{amsfonts}
\begin{document}

	\title{Signatures of fractionalization in spin liquids from interlayer thermal transport}
	\date{\today}
	\author{Yochai Werman}
	\affiliation{Department of Condensed Matter Physics, Weizmann Institute of Science, Rehovot 76100}    
    \author{Shubhayu Chatterjee}
    \affiliation{Department of Physics, Harvard University, Cambridge MA 02138, USA}
    \author{Siddhardh C. Morampudi}
        \affiliation{Department of Physics, Boston University, Boston, MA 02215, USA}
	\author{Erez Berg}
	\affiliation{Department of Condensed Matter Physics, Weizmann Institute of Science, Rehovot 76100}
	\affiliation{James Frank Institute and the Department of Physics, University of Chicago, Chicago, IL 60637, USA}
	
	\begin{abstract}
Quantum spin liquids (QSLs) are intriguing phases of matter possessing fractionalized excitations. Several quasi-two dimensional materials have been proposed as candidate QSLs, but direct evidence for fractionalization in these systems is still lacking. In this paper, we show that the inter-plane thermal conductivity in layered QSLs carries a unique signature of fractionalization. We examine several types of gapless QSL phases - a $Z_2$ QSL with either a Dirac spectrum or a spinon Fermi surface, and a $U(1)$ QSL with a Fermi surface, and consider both clean and disordered systems. In all cases, the in-plane and $c-$axis thermal conductivities have a different power law dependence on temperature, due to the different mechanisms of transport in the two directions: in the planes, the thermal current is carried by fractionalized excitations, whereas the inter-plane current is carried by integer (non-fractional) excitations. In layered $Z_2$ and $U(1)$ QSLs with a Fermi surface, and in the disordered $Z_2$ QSL with a Dirac dispersion, the $c-$axis thermal conductivity is parametrically smaller than the in-plane one, but parametrically larger than the phonon contribution at low temperatures.
\end{abstract}

	\maketitle
	
\section{Introduction} Quantum spin liquids (QSLs) are phases of matter with intrinsic topological order, which cannot be characterized by local order parameters as typically used in symmetry-breaking phases. Instead, their primary characteristic is the emergence of excitations with fractional quantum numbers~\cite{Anderson,anderson1987resonating, Lee, Balents, BalentsSavary, Kanoda}. The presence of these excitations is related to the existence of long-range entanglement in ground states of such systems~\cite{KitaevPreskill, LevinWen}. In addition, the excitations are accompanied by an emergent gauge field leading to a low-energy description in terms of gauge theories. The relevant gauge group can be discrete (e.g., $\mathbb{Z}_2$) or continuous (e.g., $U(1)$). The matter excitation spectrum may be gapped (as in a gapped $Z_2$ phase \cite{Kivelson1987,Subir1, Subir2, Subir3,SenthilFisher,MoessnerSondhi,MoessnerSondhiFradkin}) or gapless (as in a gapless $Z_2$ \cite{Kitaev,Barkeshli}  or $U(1)$ \cite{Nayak, Altshuler, Motrunich, Senthil, LeeReview, Ioffe, Nagaosa, Lee2, Polchinski} QSL). 

Several materials have been proposed as candidates for spin liquids; these three dimensional materials are often layered compounds of frustrated $2D$ lattices, such as kagome and triangular lattices. For example, members of the iridate family \cite{Jackeli, Chaloupka, Liu, Choi} have been proposed to display QSL gapless $Z_2$ behavior; the triangular organic salt $EtMe_3Sb[Pd(dmit)_2]_2$ has been proposed to have a spinon Fermi surface, while $\kappa-(ET)_2Cu_2(CN)_3$ is believed to be a gapped QSL\cite{Yamashita, Yamashita2, Yamashita3, Yamashita4}. In addition, the material Herbertsmithite is thought to be either a gapless or a small gap QSL, with its class not yet known \cite{Mendels, Shores, Jeschke, Olariu, Helton, Helton2, deVries, Bert, han2012, Pilon2013,Han2014,Fu2015}. 

The excitations of QSLs can carry fractional quantum numbers corresponding to global symmetries possessed by the system \cite{Zou1988,Kivelson1989statistics,SenthilFisher} and also possess fractional (anyonic) statistics \cite{Arovas1984,Kivelson1989statistics,Read1989,Subir1}. There have been numerous proposals to detect these fractional quantum numbers and statistics in QSL materials~\cite{Kivelson1989statistics,SenthilFisherDetecting,SenthilFisherDetecting2, Norman2009, 
barkeshli2014coherent, SCSS15, nasu2016fermionic, Sid}. The presence of fractionalization itself has primarily been deduced through a diffuse scattered intensity seen in inelastic neutron scattering experiments on various candidate spin liquids at temperatures much smaller than the relevant exchange coupling \cite{han2012, Coldea}. The absence of sharp features in the neutron scattering intensity is attributed to the presence of a multi-particle continuum \cite{Punk}. However, such broadening can also arise due to other factors such as disorder and it would be useful to have additional signatures of fractionalization. 

	In this work, we propose the inter-plane thermal conductivity $\kappa_c$ as a probe for fractionalization in a system of weakly coupled layers of two dimensional gapless QSLs \footnote{Here, we assume that the inter-layer coupling does not destabilize the layered QSL phase. This is certainly the case for a $Z_2$ QSL with a Dirac spectrum, since the inter-layer coupling is irrelevant.  For the case of a QSL with a Fermi surface, the inter-layer coupling is marginal at tree level; we assume that we are at temperatures above the (exponentially small) temperature of any possible instability}. The in-layer thermal conductivity $\kappa_{ab}$ in these materials is dominated by the low-energy fractionalized excitations pertinent to the type of QSL in question; in contrast, the thermal current between the planes must be carried by a gauge invariant excitation with integer quantum numbers.  
{This is because the emergent gauge charge carried by fractionalized excitations is conserved separately in each layer, and therefore a single spinon cannot move from one layer to the next. 
Moreover, a non-gauge invariant fractionalized excitation, such as a spinon, is highly non-local in space (it is composed of a long ``string'' of local spin operators). 
This implies that the matrix element of a local operator to transfer {\it pairs} of spinons from one layer to another decays exponentially with the spatial separation between the two spinons. Therefore, only pairs of nearby spinons can hop between adjacent layers.}  

The situation is depicted schematically in Fig.~\ref{illustration}, where a single spinon is deconfined and may propagate freely in each plane, while only pairs of spinons may hop between planes. Therefore, $\kappa_c$ in a gapless QSL is expected to be qualitatively different from the in-layer thermal conductivity, and obey a different power law at low temperatures~\footnote{A similar mechanism can provide evidence for fractionalization in the $c-$axis \emph{electrical} transport in a metallic resonating valence bond state. See: P. W. Anderson and Z. Zou, Phys. Rev. Lett. {\bf 60}, 132 (1988); N. Nagaosa, Physical Review B {\bf 52}, 10561 (1995).}. An experimental detection {of such a parametrically large anisotropy in ratio $\kappa_{ab}/\kappa_c$ at low temperatures will be a strong indication of the existence of fractionalized excitations and hence a QSL state.

\begin{table}[]
\centering
\caption{In-plane and $c$-axis thermal conductivity for several types of QSL. $Z_2$ Dirac refers to a $Z_2$ QSL with a Dirac spectrum of fermionic fractional excitations. 
$Z_2$ FS is a $Z_2$ QSL with a Fermi surface of fractional excitations. 
$U(1)$ refers to a spinon Fermi surface coupled to a $U(1)$ fluctuating gauge field. 
$\alpha = 6\Delta_A/(\pi+\Delta_A),$ with $\Delta_A$ the (dimensionless) time-reversal preserving disorder strength [see Eq.~(\ref{eq:Delta_A})]. The result for the clean $Z_2$ FS case is correct up to logarithmic factors.}
\label{table1}
\begin{tabular}{|l|l|l|l|l|}
\hline
          & \multicolumn{2}{c|}{In-plane} & \multicolumn{2}{c|}{$c$-axis} \\ \hline
          & Clean         & Disordered    & Clean         & Disordered    \\ \hline
$Z_2$ Dirac  & $T$~\cite{Durst}           & $T$           & $T^{5}$       & $T^{5-\alpha}$         \\ \hline
$Z_2$ FS &      $T^{-1}$       & $T$           &      $T^3$          & $T^2$         \\ \hline
$U(1)$    & $T^{1/3}$~\cite{Nave}     &     $T$~\cite{Nave}     & $T^{5/3}$     & $T^2$         \\ \hline
\end{tabular}
\end{table}



Our findings are summarized in Table \ref{table1}. We have considered three cases: a gapless $Z_2$ QSL with either a Dirac spectrum or a spinon Fermi surface, and $U(1)$ QSL with a spinon Fermi surface. In all cases, the in-plane and $c$-axis thermal conductivity follow qualitatively different behavior as a function of temperature, for both clean and mildly disordered systems. In all QSLs we consider, the inter-plane thermal conductivity follows a power law behavior in temperature, with an exponent which is larger than for the corresponding intra-plane behavior. Interestingly, in some cases, the exponent of the inter-plane thermal conductivity is smaller than $3$, and therefore it is parametrically larger than the phonon contribution (proportional to $T^3$) at sufficiently low temperatures. 

\begin{figure}
  \centering
    \includegraphics[width=0.4\textwidth, height=0.65\textwidth]{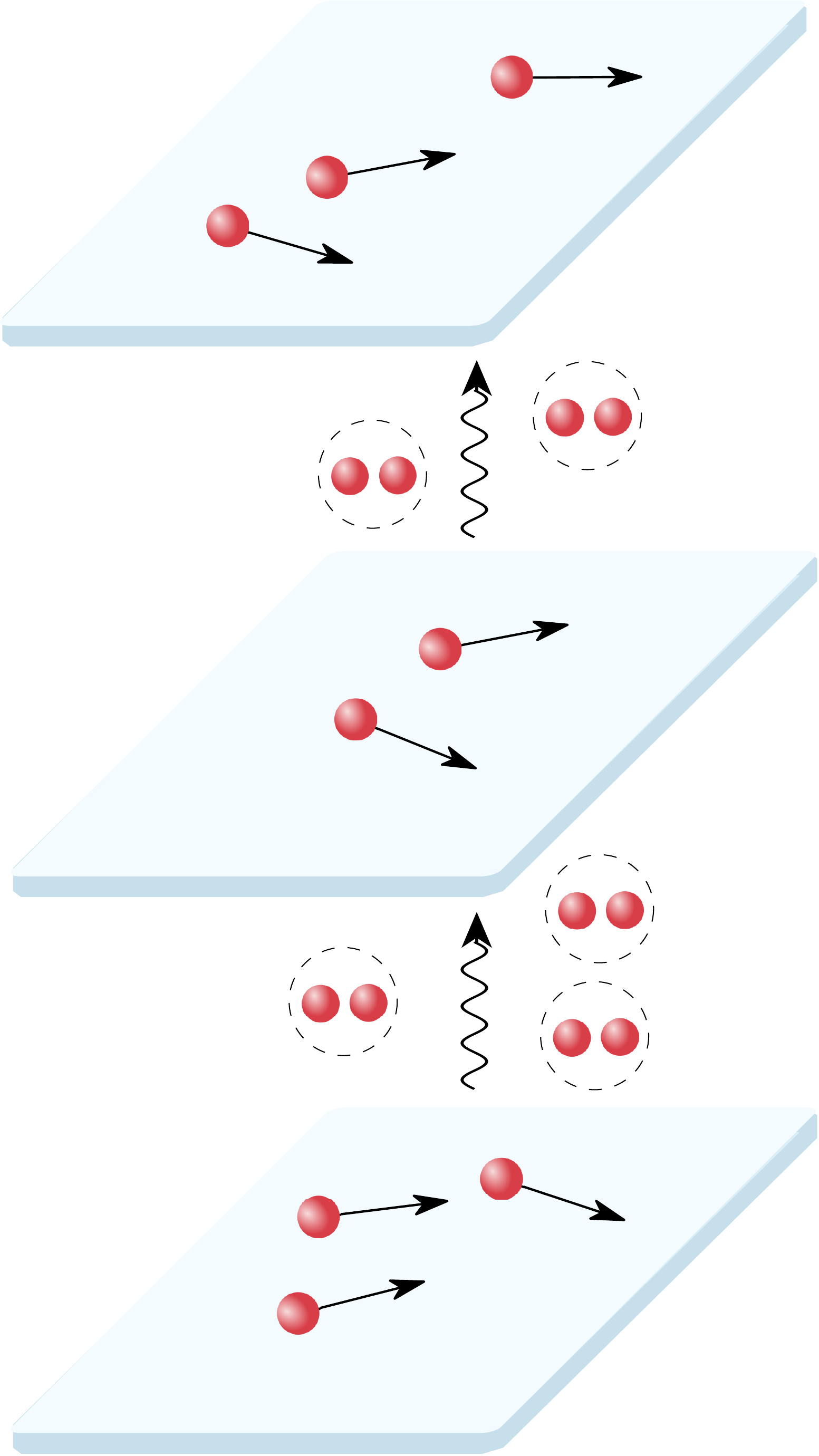}
 \caption{A schematic representation of the difference between in-plane and interplane transport. In each QSL plane, the spinons are deconfined and may travel freely. However, transport between the planes is only possible via gauge-invariant excitations, such as spinon pairs.}
\label{illustration}
\end{figure}

\section{Clean $Z_2$ Quantum Spin Liquid} We begin by considering a layered system where each layer forms a $Z_2$ QSL with gapless fermionic excitations. The fermions may either have a Dirac spectrum, or form a Fermi surface. As a concrete example of the gapless $Z_2$ QSL, one may consider the gapless phase of the Kitaev honeycomb model~\cite{Kitaev}, which consists of spin-$1/2$s interacting in an anisotropic manner on a two-dimensional hexagon lattice. We will use this model to facilitate our discussion; our conclusions are generic to any gapless $Z_2$ QSL. 

The low energy theory of the Kitaev QSL phase may be described either as two linearly dispersing Majorana fermions, or equivalently as a single complex Dirac theory. Here, we consider a three-dimensional layered generalization of the Kitaev model. The low-energy effective Hamiltonian of each layer is given by
\begin{eqnarray}
H^{Z_2}_l = \int \frac{d^2k}{(2\pi)^2} \psi^\dagger_l (\mathbf{k}) \left[ v \boldsymbol{\sigma} \cdot \mathbf{k} + m \sigma^z+\Delta \sigma^0 \right] \psi_l(\mathbf{k}),
\label{HZ2}
\end{eqnarray}
where $l$ is the layer index, and $\psi^\dagger_l(\mathbf{k}) = \left(\psi_l^{A\dagger}(\mathbf{k}), \psi_l^{B\dagger}(\mathbf{k})\right)$ is a spinor of complex fermionic spinon creation operators in layer $l$, with $A,B$ denoting the sublattice. $\boldsymbol{\sigma}$ is a vector of Pauli matrices, $\mathbf{k} = (k_x,k_y)$ is measured relative to the corner of the honeycomb lattice (the $\mathbf{K}$ point), and $v$ the Fermi velocity. {$m$ and $\Delta$ describe a mass gap and an effective chemical potential, respectively. Throughout the paper we have set $\hbar = 1$.  In Appendix~\ref{ap:z2Kitaev} we show an explicit microscopic spin Hamiltonian that leads to low-energy effective Hamiltonian in Eq.~(\ref{HZ2}). The $m$ and $\Delta$ terms arise from time reversal-breaking three spin interactions. On the honeycomb lattice, in the presence of time reversal (TR) symmetry, $m = \Delta = 0$, and the Fermi energy is at the Dirac point. Breaking time reversal symmetry~\cite{Barkeshli}, or considering generalizations of the Kitaev model to other lattices~\cite{Yang2007,Baskaran2009,Lai2011,Hermanns2014,Hermanns2015,OBrien2016}, allows for a stable Fermi surface. In all $Z_2$ QSLs we consider, the fermionic excitations (``spinons'') are gapless, which corresponds to $|\Delta|\ge |m|$. The fluxes of the $Z_2$ gauge field (``visons'') are gapped.

Although the in-plane theory is described by fractional excitations, inter-plane transport must be mediated by gauge-invariant excitations. The most relevant interlayer coupling terms which are allowed by symmetry are given by
\begin{widetext}
\begin{eqnarray}\label{couplings}
H_{\perp}^{Z_2} &=&J_\perp\sum_{\langle l,l' \rangle}\sum_{\alpha = 0,1,2,3}F_\alpha \int \frac{d^2k}{(2\pi)^2}\frac{d^2k'}{(2\pi)^2}\frac{d^2q}{(2\pi)^2} \psi_l^{\dagger}(\mathbf{k})\sigma^\alpha \psi_l (\mathbf{k+q}) \psi_{l'}^{\dagger}(\mathbf{k'})\sigma^\alpha \psi_{l'}(\mathbf{k'-q})\\
&+&J_\perp\sum_{\langle l,l' \rangle}F_4 \int \frac{d^2k}{(2\pi)^2}\frac{d^2k'}{(2\pi)^2}\frac{d^2q}{(2\pi)^2}\left[ \psi_l^{A \dagger}(\mathbf{k})\psi^{B \dagger}_l (\mathbf{-k-q}) \psi_{l'}^{A}(\mathbf{k'}) \psi_{l'}^{B}(\mathbf{k'-q})
+h.c.
\right]\nonumber,
\end{eqnarray}
\end{widetext}
where $\langle l,l' \rangle$ are neighboring layers, $\sigma^\alpha$ are the Pauli matrices (with $\sigma^0 $ the identity matrix), $J_\perp$ is the strength of the inter-plane coupling, and $F_{0,\dots,4}$ are dimensionless coupling constants. In Sec.~\ref{sec:generic} we argue that generically, $J_\perp$ is proportional to the microscopic spin-spin inter-layer interactions.

For simplicity, we will mostly focus on the case where only $F_0=F$ is nonzero. A derivation of such a coupling term from a microscopic spin-spin interaction is given in the Appendix~\ref{ap:z2Kitaev}}. We believe that the particular form of the interlayer coupling is not important;  the contribution to the thermal conductivity from other terms give the same parametric dependence on temperature. The crucial point is that the inter-plane coupling term must contain an even number of fermion operators from each layer, as a single fractional excitation may not hop from one layer to another. 

In a generic $Z_2$ QSL, there are also short-range intra-plane interactions between the fermionic spinons. However, for most of the following discussion we may ignore such interactions, as they are irrelevant in the Dirac case, and lead to a Landau Fermi liquid state with well defined quasiparticles in the Fermi surface case.

For a clean $Z_2$ QSL, whose low energy theory is described by weakly interacting fermions, the interlayer thermal conductivity may be calculated to lowest order in $J_{\perp}$ using Fermi's golden rule. We work in the basis of the eigenvalues of the in-plane Hamiltonian; we therefore revert from the sublattice ($\alpha = A,B$) to the band ($\lambda = \pm1$) basis, and consider the transformed function $F^{\lambda_1 \dots \lambda_4}_{\mathbf{k},\mathbf{k'},\mathbf{q}}$ in this basis. In the case of a Dirac spectrum, the eigenstates of the in-plane Hamiltonian are given by $a_l^{\lambda=\pm}(\mathbf{k}) = [\psi_l^{A}(\mathbf{k})\pm e^{i\phi_\mathbf{k}}\psi_l^{B}(\mathbf{k})]/\sqrt{2}$, with energy $\epsilon^\lambda_\mathbf{k} =\lambda vk$; here $\phi_\mathbf{k} = \mathrm{atan}(k_y/k_x)$. In this basis,   $F^{\lambda_1 \dots \lambda_4}_{\mathbf{k},\mathbf{k'},\mathbf{q}} =\frac{1}{4}\left[1+\lambda_1\lambda_2 e^{i(\phi_\mathbf{k}-\phi_\mathbf{k+q})}\right]\left[1+\lambda_3\lambda_4 e^{i(\phi_\mathbf{k'}-\phi_\mathbf{k'-q})}\right]$.  Energy is transported between layers by the excitation of spinon pairs; thus, if a temperature difference $\delta T$ is applied between two adjacent layers $l$ and $l'$, the rate with which energy transfer occurs, for the specific momenta $\mathbf{k}, \mathbf{k+q}, \mathbf{k'}, \mathbf{k'-q}$, is
\begin{widetext}
\begin{eqnarray}
\Gamma^E_{\mathbf{k}, \mathbf{k+q}, \mathbf{k'}, \mathbf{k'-q}} &=& \frac{2\pi}{Z} J_\perp^2 \sum_{\lambda_{1\dots 4}}|F^{\lambda_1...\lambda_4}_{\mathbf{k},\mathbf{k'},\mathbf{q}}|^2 \sum_{i_l, i_{l'};f_l, f_{l'}}\left[\exp\left(-\frac{E_{i,l}}{{T+\delta T}}-\frac{E_{i,{l'}}}{T}\right)-\exp\left(-\frac{E_{f,l}}{T+\delta T}-\frac{E_{f,{l'}}}{T}\right)\right]\\
&&\times \big| \langle f_l | \langle  f_{l'} | a^{\lambda_1 \dagger}_l(\mathbf{k})
a^{\lambda_2}_l(\mathbf{k}+\mathbf{q})a^{\lambda_3 \dagger}_{l'}(\mathbf{k'})
a^{\lambda_4}_{l'}(\mathbf{k'}-\mathbf{q}) | i_l \rangle | i_{l'} \rangle \big|^2 (E_{i,l}-E_{f,l})\delta(E_{i,l}+E_{i,{l'}}-E_{f,{l}}-E_{f,{l'}})\nonumber
\end{eqnarray}
\end{widetext}
where $\vert i_l \rangle$ ,$| f_{l} \rangle$ are the initial and final many-body states of layer $l$ (which are eigenstates of the $J_\perp=0$ Hamiltonian), with energies $E_{i,l}$ and $E_{f,l}$, respectively, and similarly for layer $l'$. $Z$ is the partition function.

The thermal conductivity is then given by (here $J^Q$ is the thermal current)
\begin{widetext}
\begin{eqnarray}\label{kappaz2clean}
\kappa_c &=& \frac{\partial J^Q}{\partial \delta T} = \int \frac{d^2k}{(2\pi)^2}\frac{d^2k'}{(2\pi)^2}\frac{d^2q}{(2\pi)^2}\frac{\partial }{\partial \delta T}\Gamma^E_{\mathbf{k}, \mathbf{k+q}, \mathbf{k'}, \mathbf{k'-q}}  \\
&=& 2\pi \frac{J_\perp^2}{T^2} \sum_{\lambda_{1\dots 4}}\int d\epsilon_1d\epsilon_2d\epsilon_3 (1-n_F(\epsilon_1))n_F(\epsilon_2)(1-n_F(\epsilon_3))n_F(\epsilon_1-\epsilon_2+\epsilon_3)
\times(\epsilon_1-\epsilon_2)^2 \nonumber\\
&\times&\int \frac{d^2k}{(2\pi)^2}\frac{d^2k'}{(2\pi)^2}\frac{d^2q}{(2\pi)^2} |F_{\mathbf{k},\mathbf{k'},\mathbf{q}}^{\lambda_1...\lambda_4}|^2\delta(\epsilon_1-\epsilon^{\lambda_1}_\mathbf{k})\delta(\epsilon_2-\epsilon^{\lambda_2}_\mathbf{k+q})\delta(\epsilon_3-\epsilon^{\lambda_3}_\mathbf{k'})\delta(\epsilon_1+\epsilon_3-\epsilon_2-\epsilon^{\lambda_4}_\mathbf{k'-q}),\nonumber
\end{eqnarray}
\end{widetext}
where $n_F(\epsilon)$ is the Fermi function. 

For the case of a $Z_2$ QSL with a Dirac spectrum, the dependence of the integral on temperature can be evaluated easily by rescaling $\epsilon\Rightarrow \epsilon/T$ and 
$\{\mathbf{k},\mathbf{k'},\mathbf{q}\}\Rightarrow \{\mathbf{k}/vT,\mathbf{k'}/vT,\mathbf{q}/vT\}$. This gives the result
\begin{equation}
\kappa_c \sim \frac{J_\perp^2}{v^6} T^5\,\,\, \mbox{(clean $Z_2$ with Dirac spectrum).}
\end{equation}

 The case of a $Z_2$ QSL with a Fermi surface corresponds to $\Delta \ne 0$ in Eq.~(\ref{HZ2}). To simplify the calculation, we set the mass term in~(\ref{HZ2}) such that $\Delta>m$ but $|\Delta-m| \ll m$. In this limit, the eigenstates of the band which crosses the Fermi energy simplify to $a(\mathbf{k}) = \psi^A(\mathbf{k})$, with a non-relativistic dispersion $\epsilon_\mathbf{k} = k^2/2m^*-\mu$, with $\mu = \Delta-m$, and $m^* = m/v^2$. The result should not depend on this choice. 

The evaluation of the integrals in Eq.~(\ref{kappaz2clean}) for the case of a Fermi surface is described in Appendix~\ref{app:integral}. After integrating over $\mathbf{k}$, $\mathbf{k}'$, and $\epsilon_{1,2,3}$, $\kappa_c$ has the form:
\begin{eqnarray}
\kappa_c&\sim J_\perp^2 T^3\int \frac{d^2k d^2k' d^2q}{(2\pi)^6}\delta(\epsilon_\mathbf{k})\delta(\epsilon_\mathbf{k+q})\delta(\epsilon_\mathbf{k'})\delta(\epsilon_\mathbf{k'-q}) \nonumber\\
&\sim J_\perp^2 \frac{\nu^4}{k_F^2} T^3 \int_0^{2k_F} dq \frac{1}{q}\frac{1}{1-q^2/4k_F^2},
\end{eqnarray}
where $\nu = m^*/2\pi$ is the density of states on the Fermi energy, and $k_F = \sqrt{2 m^* \mu}$. This integral is logarithmically divergent; this is similar to the divergence of the electronic self energy in a Fermi liquid in two dimensions~\cite{GQ1982}. As in a Fermi liquid, intralayer short-range interaction between the spinons lead to a finite spinon lifetime $\tau\propto 1/T^2$. The associated broadening of the spinon spectral function provides an infra-red cutoff for the logarithm \cite{SubirBook}, giving
\begin{eqnarray}\label{eq:cleanZ2FS1}
\kappa_c \sim J_\perp^2 \frac{\nu^4}{k_F^2} T^3 \log\left(\Lambda /T\right)\mbox{   (clean $Z_2$ with FS)}
\end{eqnarray}
with $\Lambda$ a high-energy cut-off, of the order of the Fermi energy (which is proportional to the exchange coupling between the original spins).

The in-plane thermal conductivity of the $Z_2$ QSL with a Fermi surface is given, using the Einstein relation, by $\kappa_{ab} \sim c_V v_F^2 \tau/2$, where $c_V = \pi^2 \nu T/3$ is the specific heat of the system at low temperatures, $v_F = k_F/m^*$ is the Fermi velocity, and $\tau$ is the spinon lifetime. In a perfectly clean crystal, the lifetime comes from weak short-range interaction between the spinons mediated by the gapped gauge field (assuming that Umklapp processes are available to relax the total momentum of the scattering spinons).  The lifetime is given by $\tau^{-1} \sim T^2 \log(\Lambda/T)$ as discussed earlier, and therefore we have:
\begin{eqnarray}
\kappa_{ab} \sim \left[ T  \log\left(\Lambda /T\right)\ \right]^{-1}
\end{eqnarray}

\section{Disordered $Z_2$ quantum spin liquid} As we shall now show, quenched disorder changes the low-temperature inter-plane transport in a qualitative way. The effects of disorder depend crucially on the type of disorder, which is subject to the symmetry of the problem. 
Consider, for example, the case of the honeycomb Kitaev model with time reversal symmetry. Then, disorder can take the form of a random bond strength, that translates to a random vector potential~\cite{Willans2010} in the low-energy Dirac Hamiltonian, Eq.~(\ref{HZ2}). Breaking time reversal symmetry can induce random scalar potential and mass terms, as well (see Appendix~\ref{sec:disorder_Z2} for a demonstration of how such terms arise in a disordered version of the Kitaev model).

Here, we will focus on random vector and scalar potentials; a random mass term is important at the transition between different gapped spin liquid states, a case we will not consider in the present work. The disordered part of the low-energy effective Hamiltonian in layer $l$ is given by
\begin{eqnarray}
H_{l}^{\mathrm{dis}} = \int \frac{d^2kd^2k'}{(2\pi)^4}\psi_l^\dagger(\mathbf{k})\left(\mathcal{V}_{\mathbf{k-k'}} + v\boldsymbol{\mathcal{A}_{\mathbf{k-k'}}\cdot \sigma}\right)\psi_l(\mathbf{k'}), \nonumber \\
\end{eqnarray}
where $\mathcal{V}_{\mathbf{k-k'}}$ and $\boldsymbol{\mathcal{A}_{\mathbf{k-k'}}}$ are random scalar and vector potentials, respectively. We assume that the disordered potentials in different layers are statistically independent.

First, we study the case of a Dirac QSL with time reversal symmetry, in which only a random vector potential term is allowed, $\mathcal{V}_{\mathbf{k-k'}}=0$.  
The effects of a vector potential disorder on a system with a Dirac dispersion were studied extensively in Ref.~\cite{Ludwig}, where it was shown that such a term leads to a line of fixed points, characterized by scaling exponents which depend continuously on the disorder strength. Using the methods introduced in Ref.~\cite{Ludwig}, we can find the scaling form of correlation functions at this fixed point, as described in detail in Appendix~\ref{VPdisorder}. This allows us to show that vector potential disorder results in a modification of the exponent of the thermal conductivity, which is given by
\begin{eqnarray}
\kappa_z \sim  T^{5-\alpha}\mbox{    (disordered $Z_2$ with Dirac spectrum)}, 
\end{eqnarray}
with $\alpha = 6\Delta_A/(\pi+\Delta_A)$, $\Delta_A$ being the disorder strength:
\begin{equation}
(2\pi)^2\delta(\mathbf{q}+\mathbf{q'})\Delta_A = \langle\mathcal{A}_{\mathbf{q}}\mathcal{A}_{\mathbf{q'}}\rangle_{\mathrm{dis}}, 
\label{eq:Delta_A}
\end{equation}
and the average is over disorder configurations. We consider smooth disorder, such that $\psi^\dagger\psi^\dagger$ terms (corresponding to intervalley scattering in the Majorana model) are negligible.


Next, we consider the effect of disorder on a $Z_2$ QSL with a Fermi surface, corresponding to $\Delta \ne 0$ in Eq.~(\ref{HZ2}). In this case, since time reversal symmetry is broken, both scalar and vector disorder potentials are allowed. To simplify the computation, we will neglect the vector potential in this case, and assume that the scalar potential is short range correlated in space: $\langle \mathcal{V}_{\mathbf{q}}\mathcal{V}_{\mathbf{q'}} \rangle_{\mathrm{dis}} = \delta(\mathbf{q}+\mathbf{q'})/(2\pi \nu \tau)$, where $\nu$ is the density of states at the Fermi level, and $\tau$ is the mean free time of quasi-particles at the Fermi surface. Moreover, we will again set the mass term in Eq.~(\ref{HZ2}) such that $\Delta>m$ but $|\Delta-m| \ll m$. We expect none of the qualitative aspects of the solution to depend on these choices.

In the presence of disorder, the calculation of the thermal conductivity is most conveniently done using the Luttinger prescription~\cite{Luttinger1, Luttinger2, Shastry}. The thermal conductivity is written as
\begin{equation}\label{kubo}
\kappa = \frac{-1}{T}\lim_{\omega\to 0} \frac{\Im\left[\Pi(\omega)\right]}{\omega},
\end{equation}
where $\Pi(\omega)$ is the retarded thermal current-thermal current correlation function,
\begin{equation}\label{eq:Pi}
\Pi(\omega) = \left\langle J^Q(i\omega_n) J^Q(-i\omega_n)\right\rangle|_{i\omega_n\rightarrow\omega+i\delta}.
\end{equation} 
The $c$-axis thermal current operator can be derived using the energy continuity equation: $J_c^Q(i \omega_n) = \lim_{q_c \rightarrow 0} \frac{i\omega_n h(q_c)} {q_c}$, where $h(q_c)$ is the energy density operator at wavevector $q_c$. An explicit calculation to leading order in $J_\perp$ using Eq.~(\ref{couplings}) gives (see Appendices \ref{ap:current}, \ref{sec:inter_current})
\begin{widetext}
\begin{eqnarray}\label{currentoperator}
J_c^Q (i\omega_n) 
& = & \frac{1}{32}J_\perp F_0\sum_{l,\eta=\pm 1}\eta\int \frac{d^2k}{(2\pi)^2}\frac{d^2k'}{(2\pi)^2}\frac{d^2q}{(2\pi)^2} \frac{1}{{\beta^3}}\sum_{\nu_n,\nu_m,\Omega_n}\Omega_n\nonumber\\
& \times&
\psi_l^\dagger(\mathbf{k},i\nu_n)
\psi_{l}(\mathbf{k}+\mathbf{q},i\nu_n+i\Omega_n+i\omega_n) \psi_{l+\eta}^\dagger(\mathbf{k'},i\nu_m)
\psi_{l+\eta}(\mathbf{k'}-\mathbf{q},i\nu_m-i\Omega_n)
\end{eqnarray}
\end{widetext}
Here, we have suppressed the eigenstate indices $\lambda_{1,\dots,4}$, since in the non-relativistic limit $|\Delta-m|\ll m$, the wavefunctions of states at the Fermi surface are confined to a single sublattice. Similarly, we have suppressed the eigenstate indices in $F_0,$ which is now momentum-independent. Note that, similarly to the inter-plane coupling, the thermal current operator in our model is quartic in the fermionic operators, corresponding to the fact that energy is carried between the plane by the hopping of fermion pairs.

The diagrams describing the leading-order contribution to $\kappa_c$ are shown in Fig.~\ref{fig:diagrams}. The computation is lengthy but straightforward, and we will only describe the main steps here, deferring the details to Appendix~\ref{ap:potdisorder}. We assume that the disorder is weak, such that $k_F \ell \gg 1$, where $k_F$ is the Fermi momentum and $\ell = k_F \tau / m$ is the mean free path. Under these conditions, we may use the self-consistent Born approximation~\footnote{Here, we neglect weak localization corrections, which are beyond the Born approximation. We assume that we are at not too low temperatures, such that weak localization effects are unimportant.}, equivalent to summing only non-crossed diagrams~\cite{Altland}. 


\begin{figure}
    \begin{subfigure}{0.4\textwidth}
        \centering
        \includegraphics[width=\textwidth]{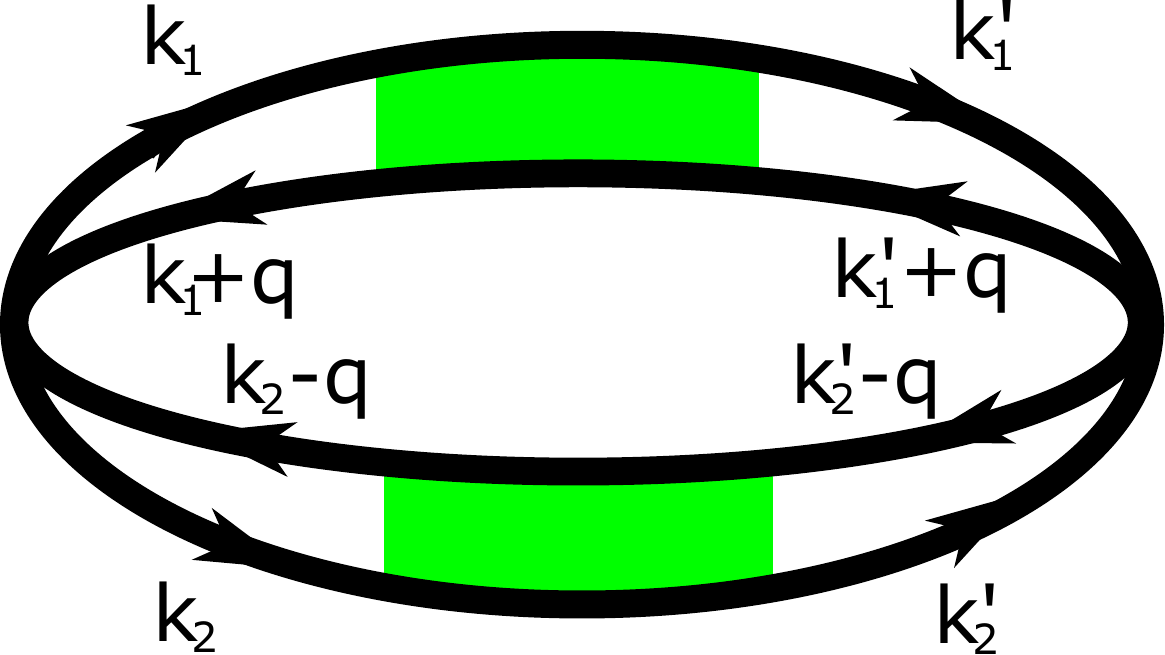}
\caption{}
    \end{subfigure}
    
    \begin{subfigure}{0.4\textwidth}
        \centering
        \includegraphics[width=\textwidth]{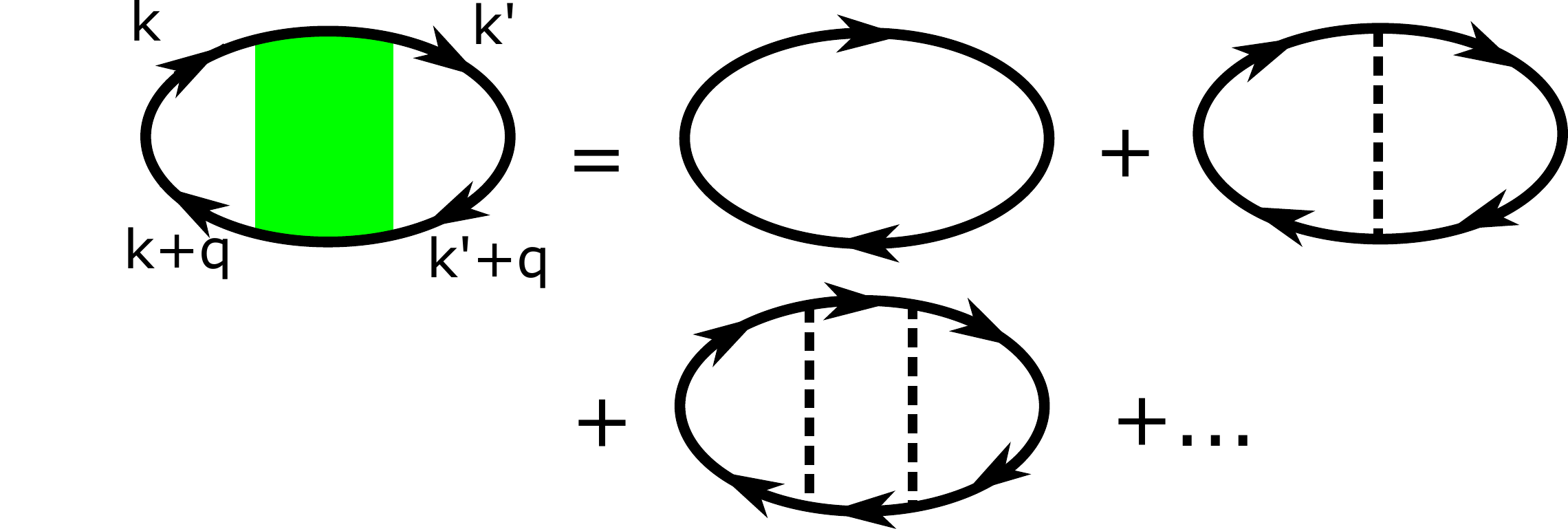}
\caption{}
    \end{subfigure}
\caption{(Color online) (a) The thermal current-thermal current correlation diagram. The Green's functions on the top, which are functions of momentum $k_1$, are related to layer $l$, while those on the bottom are from layer $l'$. Note that the current vertex consists of four Green's functions, two from each layer. (b) The disorder averaged four point correlator within each layer, $\Upsilon(\mathbf{k},\mathbf{k'},\mathbf{q})$. The black lines denote fully dressed fermion propagators, dashed lines represent the effects of disorder, the squiggly lines are the bare thermal current vertex, while the green area stands for the fully renormalized two-particle vertex. We work in the self-consistent Born approximation, applicable for $k_F \ell \gg 1$, where only ladder diagrams are taken into account. We suppress the frequency dependence for clarity.}
\label{fig:diagrams}
\end{figure}

A key object is the disorder averaged four-point correlator within a single layer, $\Upsilon$, depicted in Fig.~\ref{fig:diagrams}(b):
\begin{equation}\label{upsilon}
\begin{split}
&\Upsilon(\mathbf{k},\mathbf{k}',\mathbf{q};i\nu_n,i\nu_m)\\
& =\left\langle \psi_l^\dagger(\mathbf{k},i\nu_n)\psi_l(\mathbf{k}+\mathbf{q},i\nu_m)\psi_l^\dagger(\mathbf{k}'+\mathbf{q},i\nu_m)\psi_l(\mathbf{k}',i\nu_n)\right\rangle_\mathrm{dis}.
\end{split}
\end{equation}
The thermal current correlation function, Eq.~(\ref{eq:Pi}), is then given as a convolution of two four-point correlation functions of two adjacent layers:
\begin{eqnarray}\label{Pi}
\Pi(i\omega_n) &=& \frac{1}{64}J_\perp^2 \frac{1}{\beta^3} \sum_{\nu_n, \nu_m, \Omega_n} \int_k F_0^2 \nonumber
\\
&\times& \Omega_n^2  \Upsilon(\mathbf{k}_1, \mathbf{k}'_1, \mathbf{q}; \nu_n, \nu_n + \Omega_n) \nonumber
\\
&\times& \Upsilon(\mathbf{k}_2, \mathbf{k}'_2, -\mathbf{q}; \nu_m + \omega_n, \nu_m - \Omega_n  ).\nonumber\\
\end{eqnarray}
Here, $\int_k = \int \frac{d^2 k_1 d^2 k_2 d^2 k'_1 d^2 k'_2 d^2 q}{(2\pi)^{10}}$.

The clean, free fermion limit of this expression, with the correlator $\Upsilon(\mathbf{k}_1,\mathbf{k}_1',\mathbf{q};i\nu_n,i\nu_m)=\delta(\mathbf{k}_1-\mathbf{k}_1')G(\mathbf{k}_1,i\nu_n)G(\mathbf{k_1+q},i\nu_m)$, reproduces the Fermi golden rule calculation, Eq.~(\ref{kappaz2clean}). In the presence of disorder, the computation of $\Upsilon$ for small $q$ (such that $q\ell \ll 1$) involves a summation over a ladder series (see Appendix \ref{ap:potdisorder}); this results in
\begin{equation}
\label{Ups}
\begin{split}
&\Upsilon(\mathbf{k}, \mathbf{k}',\mathbf{q}; i\nu_n, i \nu_m) \approx G(\mathbf{k}, i\nu_n) G(\mathbf{k + q}, i\nu_m)\\
&\times \left\{\delta(\mathbf{k}-\mathbf{k}') + 1/(2\pi\nu\tau^2)\right.\\
&\times \left. [{|\nu_n - \nu_m|+D\mathbf{q}^2}]^{-1}G(\mathbf{k}', i\nu_n)G(\mathbf{k'+q}, i\nu_m)\right\}
\end{split}
\end{equation}
where 
$D$ the diffusion constant $D =v^2\tau/2$, with $\tau$ the disorder induced single particle lifetime, and
\begin{equation} 
G(\mathbf{k}, i\nu_n) = \frac{1}{i\nu_n-\epsilon_\mathbf{k} + i \mathrm{sgn}(\nu_n)/2\tau}.
\end{equation} 
Note the appearance of the diffusion kernel in Eq.~(\ref{Ups}); this is related to the diffusive behavior of the dynamical charge correlation function in a disordered system. 

The computation of the sums in Eq.~(\ref{Pi}) is described in Appendix~\ref{ap:potdisorder}. The dominant contribution comes from low frequencies and momenta, where the four-point correlator takes the form (\ref{Ups}). At low temperatures, $T < 1/\tau$, the result is

\begin{eqnarray}
\kappa_c \sim J_\perp^2\frac{\nu^2}{D}T^2 (\mbox{disordered $Z_2$ with Fermi surface}).
\label{eq:Z2_FS}
\end{eqnarray}
At higher temperatures, $T \gtrsim 1/\tau$, $\kappa_c$ crosses over to the clean form, Eq.~(\ref{eq:cleanZ2FS1}).  Eq.~(\ref{eq:Z2_FS}) can also be derived from scaling arguments, assuming that the intra-plane density-density correlation function has a diffusion form; see Appendix~\ref{sec:disordered_FS}.

In Appendix \ref{ap:pairhopping}, we show that the $F_4$ pair hopping inter-layer term results in the same power law, $\kappa_c \sim T^2$.

\section{$U(1)$ Quantum Spin Liquid-} We further study the case of a layered $U(1)$ QSL with a spinon Fermi surface. In addition to the fermionic spinons, 
there exist gapless gauge field photons, which also contribute to transport. The low energy sector of each layer is described by the Lagrangian density~\cite{Motrunich, Senthil, LeeReview, Ioffe, Nagaosa, Lee2, Polchinski}
\begin{eqnarray}
L_l &=&\sum_{\sigma=\uparrow,\downarrow} \psi_{l,\sigma}^\dagger \left(\partial_t-ia_0-\mu\right)\psi_{l,\sigma} \nonumber \\
&+& \frac{1}{2m}\psi_{l,\sigma}^\dagger \left(-i\nabla-\mathbf{a}\right)^2\psi_{l,\sigma}. 
\end{eqnarray}
where $\psi^\dagger_{l,\sigma}$ creates a spinon at layer $l$ with spin $\sigma$, $(a_0, \mathbf{a})$ is the $U(1)$ gauge field, $\mu$ is a chemical potential that sets the size of the spinon Fermi surface and $m$ is the spinon effective mass. 
A ``Maxwell'' term for $a_\nu$, $\frac{1}{2g}\sum_{\nu\lambda} f^{\nu\lambda} f_{\nu\lambda}$ where $g$ is a coupling constant and $f_{\nu\lambda} = \partial_{\nu} a_{\lambda} - \partial_{\lambda} a_{\nu}$, is also allowed by symmetry; however, it gives rise to sub-leading contributions at low momenta and frequencies, and hence we will drop it in the following. 

Under the random phase approximation (RPA), the clean system is described by a strong-coupling fixed point, with the retarded gauge boson and spinon propagators ($D^R(\mathbf{q}, \omega)$ and $G^R(\mathbf{k}, \omega)$, respectively) given by
\begin{eqnarray}\label{eq:propagators}
D^R_{\alpha \beta}(\mathbf{q}, \omega) &=& P_{\alpha \beta}(\mathbf{q})\left[-i\gamma\frac{\omega}{q}+\chi q^2\right]^{-1}\\
G^R(\mathbf{k}, \omega) &=& \left[c(-i\omega)^{2/3}-\xi_\mathbf{k}\right]^{-1},\mbox{ (clean $U(1)$ QSL)}\nonumber
\end{eqnarray}
with $\xi_\mathbf{k} = k^2/m-\mu$ the spinon energy, $\gamma = k_F/\pi$, $\chi=1/(12\pi m)$, $c = (k_F/m)\chi^{-2/3}k_0^{-1/3}$ and $ P_{\alpha \beta}(\mathbf{q}) = \delta_{\alpha \beta} - q_\alpha q_\beta/q^2$, with $k_0$ of the order of $k_F = \sqrt{2m\mu}$, the Fermi momentum. 
The use of the RPA has been formally justified in a large-$N$ expansion, where $N$ is the number of fermion flavors~\cite{Polchinski}, but this has been shown to be problematic~\cite{SSLee}. Additional expansion parameters have been proposed, that essentially reproduce the RPA results~\cite{Mross, Dalidovich}.  We shall use the RPA approximation, assuming it is pertinent to at least some area in parameter space.

In a layered $U(1)$ QSL, heat may be transferred between the layers both by spinon and photon excitations. The most relevant inter-layer interaction term of each sector is given by
\begin{eqnarray}\label{U1couplings}
H_{\perp}^{U(1)} &=& J^{sp}_\perp\sum_{\langle l,l' \rangle} \int \frac{d^2k}{(2\pi)^2}\frac{d^2k'}{(2\pi)^2}\frac{d^2q}{(2\pi)^2} F^{\sigma_{1,\dots,4}}_{sp}({\mathbf{k},\mathbf{k'},\mathbf{q}})\nonumber\\
 &\times&\psi_{l \sigma_1}^\dagger(\mathbf{k}) \psi_{l\sigma_2} (\mathbf{k+q}) \psi_{l' \sigma_3}^\dagger(\mathbf{k'}) \psi_{l' \sigma_4}(\mathbf{k'-q})\nonumber\\
&+& J^{ph}_\perp\int \frac{d^2k}{(2\pi)^2} k^2 F_{ph}(\mathbf{k}) \mathbf{a}^T_l(\mathbf{k}) \mathbf{a}^T_{l'} (\mathbf{k}) .
\end{eqnarray}
where $ \mathbf{a}^T$ is the transverse part of the gauge field. The coupling functions $F_{sp}$ and $F_{ph}$ depend on the spatial structure of the inter-layer coupling; their explicit form is unimportant. In real space, the gauge invariant term $\nabla\times \mathbf{a}$ is related the chirality of the underlying spin degrees of freedom \cite{Lee2}, and therefore the $J^{ph}_\perp$ term corresponds to an interaction between the chiralities of the spin textures in the two layers. Micropically, this term may be small compared to $J^{sp}_\perp$, since it is of higher order in the inter-plane Heisenberg exchange coupling. However, as we shall see below, in a clean case, it gives a dominant contribution to $\kappa_c$ at asymptotically low temperatures.

The calculation of the spinon-mediated inter-plane thermal conductivity proceeds in a similar fashion as in the $Z_2$ QSL case. $\kappa_c$ is given by a similar expression to Eq.~(\ref{kappa}) (with the replacement $F\rightarrow F_{sp}$). It is given by
\begin{equation}
\kappa_{c,sp}\approx  (J^{sp}_\perp)^2 \frac{\nu^2}{v^2} T^3\log\left(\frac{T}{(W/c)^{3/2}}\right).
\end{equation} 
where $W$ is an appropriate UV cut-off (see Appendix \ref{ap:U1}).

However, in the clean case, the dominant source of low-$T$ thermal transport turns out to be the exchange of gauge fluctuations; this contribution may also be calculated by the Kubo formula, and is given by (see Appendix \ref{ap:U1} for details)
\begin{eqnarray}
\kappa_{c,ph}&=&\frac{(J^{ph}_{\perp})^2}{T}\int \frac{d^2k}{(2\pi)^2} k^4 \int_0^\infty d\epsilon A_{ph}^2(\mathbf{k},\epsilon)\epsilon^2\partial_\epsilon n_B(\epsilon)\nonumber\\
 &\sim& (J^{ph}_{\perp})^2\gamma^{2/3}\chi^{4/3}T^{5/3}\mbox{  (Clean $U(1)$ QSL)}.
\end{eqnarray}
Here $A_{ph}(\mathbf{k}, \epsilon)= -2\Im D^R(\mathbf{k}, \epsilon) = \gamma\frac{|\epsilon|k}{\chi^2 k^6+\gamma^2\epsilon^2}$ is the photon spectral function. 

Thus, at sufficiently low temperature, $\kappa_{c,ph} \gg \kappa_{c,sp}$. Note that the thermal conductivity can be written as $\kappa_{c,ph} \sim T^{2 - 1/z}$, where $z=3$ is the dynamical critical exponent of the fixed point described by RPA.

The introduction of disorder to the $U(1)$ theory is likely to destabilize the $z=3$ fixed point, leading instead to diffusive behavior, similar to that of a disordered Fermi liquid. In the RPA approximation, the propagators of the disordered theory are given by~\cite{Galitski}
\begin{eqnarray}\label{eq:propagators2}
D^R_{\alpha \beta}(\mathbf{q}, \omega) &=& P_{\alpha \beta}(\mathbf{q}) \left[-i{\omega}+Dq^2\right]^{-1}\\
G^R(\mathbf{k},\omega) &=& \left[\omega-\xi_\mathbf{k}+i/(2\tau)\right]^{-1} \mbox{ (disordered $U(1)$ QSL)}\nonumber
\end{eqnarray}
with $D$ a diffusion constant and $\tau$ the disorder-induced finite lifetime. 
The calculation of the c-axis thermal conductivity is then similar to the disordered $Z_2$ QSL case. Inserting Eq.~(\ref{eq:propagators2}) in the Kubo formula for the c-axis conductivity leads to
\begin{eqnarray}
\kappa_c\propto T^2 \mbox{(disordered $U(1)$ QSL)}
\end{eqnarray}
for both spinon and photon contribution.  

\section{Experimental considerations}
In this section, we discuss possible experimental candidate systems where thermal conductivity provides a gateway to observing QSL physics. In order to observe the magnetic contribution to the inter-layer thermal conductivity, one has to be able to separate it from the phonon contribution. Since the phonon contribution scales as $T^3$, the magnetic contribution in certain QSLs dominates at sufficiently low temperatures. This happens in QSLs with a disordered spinon Fermi surface and in strongly disordered Dirac QSLs (see Table~\ref{table1}). Below, provide a rough order-of-magnitude estimate for the temperature $T_*$ at which the magnetic contribution exceeds the phonon one, as a function of system parameters (such as the strength of the inter-plane coupling, the Debye temperature, and the disorder strength). As we elaborate below, this estimate indicates that at least in some material candidates, the crossover to magnetically dominated thermal transport may occur at accessible temperatures. 



We base our estimate of $T_*$ on the case of a QSL with a spinon FS, whose magnetic c-axis thermal conductivity is given by Eq.~(\ref{eq:Z2_FS}). 
We set the unit of length to be the lattice spacing $a$, and estimate
$\nu\sim1/J$, $D = \frac{1}{2}v_{F}\ell_{sp} \sim \frac{1}{2}  J\ell_{sp}$, where $J$ is the in-plane exchange coupling, and $\ell_{sp}$ is the spinon mean-free path in the plane. This gives
\begin{equation}
\kappa_{sp}\sim\frac{2J_{\perp}^{2}T^{2}}{J^{3}\ell_{sp}}.\label{eq:kappa_sp}
\end{equation}
Next, we estimate the contribution of the phonons. The acoustic phonon
specific heat is $c_{V}\sim(T/\Theta_{D})^{3}$, where $\Theta_D$ is the Debye frequency. The (three-dimensional) phonon diffusivity
is $D_{ph}= \frac{1}{3} c_s\ell_{ph}\sim \frac{1}{3} \Theta_{D}\ell_{ph}$, where $c_s$ is the sound velocity. Therefore, by the Einstein relation,
\begin{equation}
\kappa_{ph}\sim\frac{T^{3}}{3\Theta_{D}^{2}}\ell_{ph}.
\label{eq:kappa_ph}
\end{equation}
The temperature below which the spinon contribution to the thremal conductivity becomes larger than the phonon contribution is given by equating (\ref{eq:kappa_sp}) to (\ref{eq:kappa_ph}). The result is
\begin{equation}
T_{*}=\frac{6J_{\perp}^{2}\Theta_{D}^{2}}{J^{3}\ell_{ph}\ell_{sp}}.
\label{eq:Tstar}
\end{equation}
Eq.~(\ref{eq:Tstar}) highlights the parameters that control $T_*$: $T_*$ is higher the stronger the disorder, the higher is $\Theta_D$, and the smaller is $J$. [Note that Eq.~(\ref{eq:kappa_sp}) is only valid for $T\ll J$; therefore, $T_*$ in Eq.~(\ref{eq:Tstar}) cannot exceed $J$]. 

As an illustrative example, we roughly estimate the crossover temperature $T_*$ for kapellasite, a kagome gapless QSL candidate~\cite{Wills2012}. This is a polymorph of Herbertsmithite; however, the in-plane exchange coupling is about an order of magnitude smaller. 
The exchange couplings of kapellasite have been estimated from
from first-principle calculations \cite{Jeschke2013}:
$J\approx10\mathrm{K}$, $J_{\perp}\approx0.5\mathrm{K}$. We assume that kapellasite has a spinon Fermi surface, and that the Debye temperature is $\Theta_{D}\sim 300\mathrm{K}$. 
The mean free paths of the spinons and the phonons are not known. However, disorder in the planes is believed to be substantial. To get a rough estimate of the order of magnitude of $T_*$, let us assume a strongly disordered sample, such that $\ell_{sp}=20a$ and $\ell_{ph}=200a$. This gives
\begin{equation}
T_*\approx 6 \frac{0.5^{2}\cdot300^{2}}{10^{3}\cdot 20 \cdot 200} \approx 35 \, \mathrm{mK}.
\end{equation}
kapellasite does not order magnetically at least down to $20 \, \mathrm{mK}$ \cite{Wills2012}. Thus, for sufficiently strong disorder, we get that the crossover temperature is within experimental reach. 

Let us discuss other QSL candidate materials where the spinon contribution to $\kappa_c$ may be measurable. A promising candidate material is the recently discovered 2d spin-orbit coupled iridate H$_3$LiIr$_2$O$_6$, which has been observed to be paramagnetic to very low temperatures, and hosts gapless excitations \cite{Takagi,Slagle2017}. Compared to other similar compounds like Na$_2$LiO$_3$ and Li$_2$IrO$_3$ (which order at low temperatures), in H$_3$LiIr$_2$O$_6$ the interlayer distance is smaller due to replacement of Li by smaller H atoms in between layers, which increases $J_\perp$. Further, the in-plane bond length is also larger which reduces the scale of in-plane exchange interactions $J$. As per Eq.~(\ref{eq:Tstar}), both these factors are conducive to a larger crossover temperature $T_*$ where the magnetic contribution becomes large.


Other candidate materials are magnetic insulators with strong spin-orbit coupling, where the Kitaev interaction is the dominant term. Some of these materials, like $\alpha-$RuCl$_3$, are believed to be proximate to a QSL phase \cite{Nagler1}. 
Further, the magnetic order can be suppressed by doping, making such materials an interesting playground for observing spin-liquid physics~\cite{Nagler2}, although the nature of the field-induced QSL phase is still unclear. 

In the layered organic insulators~\cite{Kanoda}, the inter-plane exchange coupling is estimated to be three orders of magnitude below the intra-plane coupling~\footnote{M. Yamashita, private communication.}, and therefore it is likely that phonons dominate the c-axis thermal transport at accessible temperatures. Herbertsmithite~\cite{Kanoda} is believed to have a gapped QSL ground state~\cite{Fu2015}, although the spin gap seems to be quite small ($\Delta_{gap} \lesssim 10 K $~\cite{Fu2015,Han2016}). An applied magnetic field can induce a finite spinon density of states at zero energy, opening the way to measure the spinon contribution to $\kappa_c$. However, the in-plane exchange coupling $J$ is about an order of magnitude larger larger than in kapellasite, while the ratio $J_\perp/J$ is comparable in the two systems~\cite{Jeschke2013}. Therefore, we expect $T_*$ in Herbertsmithite to be smaller than in kapellasite.



Finally, we discuss a few techniques can be used to isolate the magnetic contribution to the thermal conductivity from that of phonons. 


(i) In gapless spin liquid candidates where the magnetic contribution is a power law of the form $T^\theta$ with $\theta < 3$, one can isolate the magnetic contribution from the phononic one (which scales as $T^3$), since the magnetic contribution is dominant at low sufficiently low temperature. Plotting a curve of $\kappa/T^{\theta}$ vs. $T^{3-\theta}$, the slope of the curve gives us the phonon contribution, while the intercept gives us the magnetic contribution to the thermal conductivity. This is possible as long as the sample temperature is not much higher than $T_*$. 

(ii) In addition, in some materials an applied magnetic field may be used to establish long range order, suppressing the spinon contribution to the thermal conductivity, while weakly affecting the phonon contribution. Contrasting the measurements of the $c$-axis thermal conductivity in the presence and absence of such a field may enable us to isolate the spinon contribution. \\

\section{Conclusions} We have studied the thermal conductivity in layered, gapless QSLs. The key observation is that the mechanisms of in-plane and out-of-plane thermal transport are qualitatively different: the former is carried by fractionalized excitations, while the latter is carried by gauge-neutral, non-fractionalized excitations. Thus, in all the cases we have studied, $\kappa_{ab}$ and $\kappa_c$ follow different power law dependences at low temperature; in particular, the anisotropy $\kappa_{ab}/ \kappa_c$ diverges in the limit $T\rightarrow 0$. This property is a clear hallmark of a fractionalized, layered system. A large number of layered QSL candidates have been proposed in the last few years, and inter-plane thermal conductivity can serve as an unambiguous probe for fractionalization in these experimental candidates.

\acknowledgements

We thank S. Choi, J. Chalker, K. Michaeli, S. Kivelson, T. Senthil, S. Trebst, and M. Yamashita for useful discussions. E. B. and Y. W. were supported in part by the European
Research Council (ERC) under the European
Unions Horizon 2020 research and innovation programme
(grant agreement No 639172), and by the Deutsche Forschungsgemeinschaft (CRC 183). SC acknowledges support from the Harvard-GSAS Merit Fellowship. SM acknowledges support from the NSF through grant no. PHY-1656234.



\bibliography{paper}

\appendix
\widetext

\section{Layered Kitaev honeycomb model} 
\label{ap:z2Kitaev}

\subsection{Intra-layer Hamiltonian}
\label{sec:intra}

We model the layered $Z_2$ QSL system as layers of the Kitaev honeycomb model, coupled by a weak inter-layer interaction.
The Kitaev honeycomb model \cite{Kitaev} is an exactly solvable model of interacting spin-$1/2$s. It is composed of a honeycomb lattice of spins interacting via direction-dependent exchange interactions, 
\begin{eqnarray}
H_0 &= &-J\sum_{\langle j,k\rangle}S_j^{\alpha_{jk}}S_k^{\alpha_{jk}},
\end{eqnarray}
where $j,k$ are nearest neighbors on the hexago lattice, and $S^\alpha_{jk}$ are the $x$, $y$, or $z$ component of the spin operator, depending on the type of link between $j$ and $k$. The links are denoted $x$, $y$, or $z$, based on their orientation, as shown in Fig. \ref{hexagon}. Each of the spins is represented in terms
of Majorana fermions $b_i^x, b_i^y, b_i^z, c_i$ as $S_i^\alpha = ib_i^\alpha c_i$; however, the representation in terms of these fermions spans a larger Fock space, and must be restricted to the physical Hilbert space of the spins by the gauge 
$ D_i =b_i^x b_i^y b_i^z c_i = 1$. 
On each $\alpha$-direction link, $u^\alpha_{ij} = ib_i^\alpha b_j^\alpha$ is conserved, and a theorem by Lieb~\cite{Lieb1994} guarantees that the ground state is in the sector where it is possible to set $u^\alpha_{ij} = 1$ (the flux-free sector). Thus, the ground state manifold is described by the free Majorana Hamiltonian
\begin{eqnarray}
H_0 = \frac{1}{2}iJ\sum_{i,\delta} A_iB_{i+\delta},
\end{eqnarray}
with $A_i$ and $B_{i+\delta}$ the $c_i$ Majorana on the $A$ and $B$ sublattice, respectively. The $\delta$s are the three vectors connecting the even and odd sublattices, as shown in Fig.~\ref{hexagon}. 

In order to probe the $Z_2$ QSL with a Fermi surface, we consider further two specific time-reversal breaking terms, which result in a Fermi surface 
without creating vison excitations that would take us out of the ground state manifold:

\begin{eqnarray}
H_{TRB} &=& J^1_{TRB}\sum_{plaquettes}\left[S_1^z S_2^y S_3^x-S_2^x S_3^z S_4^y+S_3^y S_4^x S_5^z-S_4^z S_5^y S_6^x+S_5^x S_6^z S_1^y-S_6^y S_1^x S_2^z\right]\nonumber\\
&+&J^2_{TRB}\sum_{plaquettes}\left[S_1^z S_2^y S_3^x+S_2^x S_3^z S_4^y+S_3^y S_4^x S_5^z+S_4^z S_5^y S_6^x+S_5^x S_6^z S_1^y+S_6^y S_1^x S_2^z\right].
\end{eqnarray}

Here, the sites labeled $1...6$ on each plaquette are shown in Fig.~\ref{hexagon}.

In the ground state manifold, these terms are given by
\begin{eqnarray}\label{eq:nearestneighbors}
&=&-iJ^1_{TRB}\sum_{plaquettes}\left[A_1A_3 - B_2B_4 + A_3A_5 - B_4B_6 + 
A_5A_1 - B_6B_2\right]\nonumber\\
&-&iJ^2_{TRB}\sum_{plaquettes}\left[A_1A_3 + B_2B_4 + A_3A_5+ B_4B_6 + 
A_5A_1 + B_6B_2\right].
\end{eqnarray} 

We comment that on different lattices, one can also obtain a QSL with a spinon Fermi surface even in presence of TRS~\cite{Yang2007,Baskaran2009,Lai2011,Hermanns2014,Hermanns2015,OBrien2016}.  

Lastly, we consider also the effect of disorder in the system via a term $H_{dis}$, the exact form of which will be given later. Thus, the spin liquid we consider is described by
\begin{eqnarray}\label{Hamiltonian}
H = H_0+H_{TRB}+H_{dis},
\end{eqnarray}

with $H_{TRB}=0$ in the TR invariant case.

\subsection{Inter-layer coupling}
\label{sec:inter}

The most relevant interlayer coupling terms are those that leave each layer in its ground state; that is, the interlayer coupling must commute with all the $u_{ij}^\alpha$s. A general form of such a tunneling term in the language of the original spins will consist of spin operators from one layer coupled to spin operators from another. In order to maintain exact solvability, we consider the interlayer coupling term
\begin{eqnarray}
H_\perp &=& J_\perp\sum_{l,l'=l\pm 1}\sum_{plaquettes}\left[S_{1,l}^z S_{2,l}^y S_{3,l}^x-S_{2,l}^x S_{3,l}^z S_{4,l}^y+S_{3,l}^y S_{4,l}^x S_{5,l}^z-S_{4,l}^z S_{5,l}^y S_{6,l}^x+S_{5,l}^x S_{6,l}^z S_{1,l}^y-S_{6,l}^y S_{1,l}^x S_{2,l}^z\right]\nonumber\\
&&\times\left[S_{1,l'}^z S_{2,l'}^y S_{3,l'}^x-S_{2,l'}^x S_{3,l'}^z S_{4,l'}^y+S_{3,l'}^y S_{4,l'}^x S_{5,l'}^z-S_{4,l'}^z S_{5,l'}^y S_{6,l'}^x+S_{5,l'}^x S_{6,l'}^z S_{1,l'}^y-S_{6,l'}^y S_{1,l'}^x S_{2,l'}^z\right].
\label{eq:inter}
\end{eqnarray}

In the Majorana representation this term is given by (again setting $\hat{u}_{j,\delta} =1$)
\begin{eqnarray}
H_\perp &=& J_\perp\sum_{l,l'=l\pm 1} \sum_{plaquettes} \left[A_{1l}A_{3,l} - B_{2,l}B_{4,l} + A_{3,l}A_{5,l} - B_{4,l}B_{6,l} + 
A_{5,l}A_{1,l} - B_{6,l}B_{2,l}\right]\nonumber\\
&\times&\left[A_{1,l'}A_{3,l'} - B_{2,l'}B_{4,l'} + A_{3,l'}A_{5,l'} - B_{4,l'}B_{6,l'} + 
A_{5,l'}A_{1,l'} - B_{6,l'}B_{2,l'}\right].
\label{eq:interplane}
\end{eqnarray}
We have chosen this term as it allows for simple calculations, and in particular it reduces to a ``density-density'' interlayer interactions [the $F_0$ term in Eq.~(\ref{couplings})] in the continuum limit. 
We expect the exact form of the coupling term to be unimportant; the important point is that it must have at least two spinon operators from each layer. This is because only a spinon pair excitation can be transferred between different layers, and this requires at least two spinon operators from each. We will comment on the case of a more generic form of the inter-layer Hamiltonian, having on spin operator in each layer, in Sec.~\ref{sec:generic} below.

\subsection{Continuum Hamiltonian}

Using $f_\mathbf{k} = \sum_\delta e^{i\mathbf{k}\cdot\mathbf{\delta}},$ $g_\mathbf{k} = \sum_i\sin(\mathbf{k}\cdot\mathbf{n}_i)$, $
\psi_l^A(\mathbf{k})= e^{i\pi/4}\sum_j A_j^l e^{i\mathbf{k}\cdot\mathbf{R}_j},$ 
$\psi_l^B(\mathbf{k}) = e^{-i\pi/4}\sum_j B_j^l e^{i\mathbf{k}\cdot\mathbf{R}_j}$,
$\Delta_\mathbf{k} = 4J^1_{TRB}g_\mathbf{k},$
and $m_\mathbf{k} = 4J^2_{TRB}g_\mathbf{k}$ (where the vectors $\mathbf{n}_{i=1,2,3}$ are defined in Fig.~\ref{hexagon}), the disorder free Hamiltonian of each layer is given by

\begin{eqnarray}\label{hperpz2}
&&H^l\equiv H_0^l+H_{TRB}^l =\frac{1}{4}\int \frac{d^2k}{(2\pi)^2}\left( \begin{array}{cc}
\psi^{A\dagger}_l(\mathbf{k}) & \psi^{B\dagger}_l(\mathbf{k})  \end{array} \right) 
\left( \begin{array}{cc}
\Delta_\mathbf{k} +m_\mathbf{k} & Jf_\mathbf{k} \nonumber\\
Jf^*_\mathbf{k} & \Delta_\mathbf{k}-m_\mathbf{k}
\end{array} \right) 
\left( \begin{array}{cc}
\psi^A_l(\mathbf{k}) \nonumber \\
\psi^B_l(\mathbf{k})
\end{array} \right)
\end{eqnarray}
\begin{eqnarray}\label{hperpz21}
H_\perp &=& \frac{1}{2}{J_\perp}\sum_{l,l' = l\pm1}\int \frac{d^2k}{(2\pi)^2}\frac{d^2k'}{(2\pi)^2}\frac{d^2q}{(2\pi)^2} g_{\mathbf{k}+\mathbf{k'}}\left[\psi^{A\dagger}_l(\mathbf{k})
\psi^{A}_l({\mathbf{k}+\mathbf{q}})+\psi^{B\dagger}_l(\mathbf{k})
\psi^{B}_l({\mathbf{k}+\mathbf{q}})\right]\nonumber\\
&\times&\left[\psi^{A\dagger}_{l'}(\mathbf{k'})
\psi^{A}_{l'}({\mathbf{k'}-\mathbf{q}})+\psi^{B\dagger}_{l'}(\mathbf{k'})
\psi^{B}_{l'}({\mathbf{k'}-\mathbf{q}})\right]
\end{eqnarray}


\begin{figure}
\centering
\includegraphics[width=0.5\textwidth]{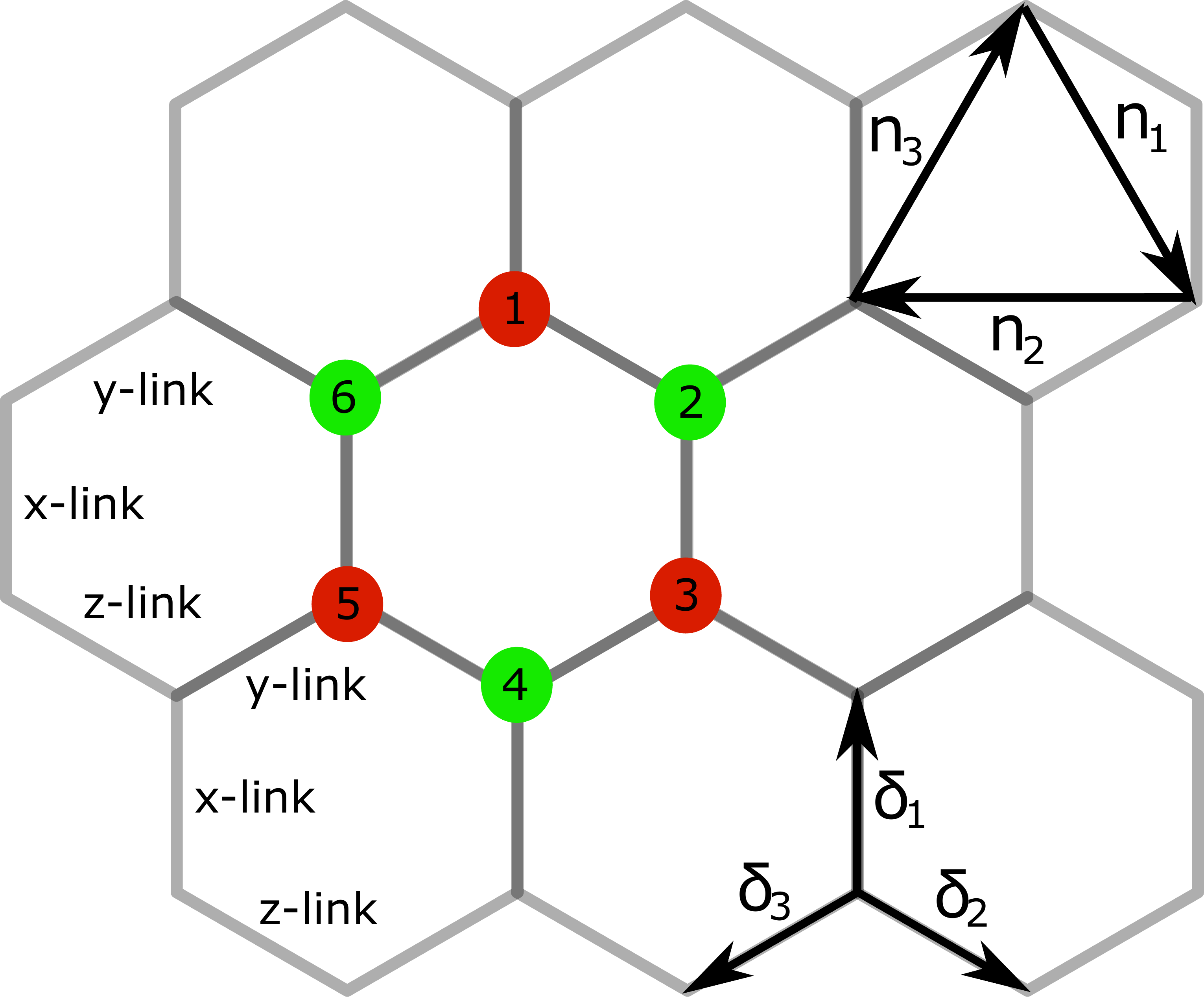}
  \centering
\caption{(Color online) The honeycomb lattice. Each unit cell is composed of an $A$ (blue) and $B$ (red) atom. Each atom in the $A$ sublattice is connected to three $B$ atoms via the vectors $\delta$, and to six A atoms via the $\mathbf{n}$ vectors. The interaction between neighbors is determined by the link they share. On each plaquette, the sites are labeled as shown.}
\label{hexagon}
    \end{figure}

The low energy theory is centered near the Dirac points $\mathbf{K,K'} = \frac{2\pi}{3}\left(\pm\frac{1}{\sqrt{3}},1\right)$. Near these points, $Jf_\mathbf{k}\approx v(kx\pm ik_y)$, with $v = 3J/2$. The low energy in-plane Hamiltonian can thus be written as two Majorana theories with a Dirac dispersion. It is convenient to consider an equivalent system, of complex fermions which reside only on half the Brillouin zone; this makes use of the equivalence $\psi^A(\mathbf{k}) = \psi^{A\dagger}({\mathbf{-k}})$. The low energy theory of this system is given by a single Dirac cone centered at $\mathbf{K}$, and its in-plane Hamiltonian is given by $H^l = \int\frac{d^2k}{(2\pi)^2}\psi^{l\dagger}_\mathbf{k}H(\mathbf{k})\psi^{l}_\mathbf{k}$, with

\begin{eqnarray}
H(\mathbf{k}) = v\boldsymbol{\sigma \cdot k}+m\sigma_z+\Delta.
\end{eqnarray}
Here $\psi^{\dagger l}_\mathbf{k} = (\psi_l^{A\dagger}(\mathbf{k}) ,\psi^{B\dagger}_l(\mathbf{k}))$ is a spinor of complex fermions, $\boldsymbol{\sigma} = (\sigma_x,\sigma_y)$ a vector of Pauli matrices, $\Delta = \Delta_{\mathbf{k} = \mathbf{K}}$, and $m = m_{\mathbf{k} = \mathbf{K}}$. This is Eq.~(\ref{HZ2}) in the main text.

In terms of the continuum theory, 
the inter-plane term in Eq.~(\ref{eq:interplane}) is given by
(neglecting the variation of $\Delta_\mathbf{k}$ around the Dirac points)
\begin{eqnarray}
H_\perp &=& \frac{1}{2}{J_\perp}\sum_{l,l' = l\pm1}\int \frac{d^2k}{(2\pi)^2}\frac{d^2k'}{(2\pi)^2}\frac{d^2q}{(2\pi)^2}\\
&\times& \left\{g_{\mathbf{k+k'}}\left[\psi_l^{A\dagger}(\mathbf{k})\psi^A_l (\mathbf{k+q})+\psi_l^{B\dagger}(\mathbf{k})\psi^B_l (\mathbf{k+q})\right] \left[\psi_{l'}^{A\dagger}(\mathbf{k'}) \psi_{l'}^A(\mathbf{k'-q})
+\psi_{l'}^{B\dagger}(\mathbf{k'}) \psi_{l'}^B(\mathbf{k'-q})\right] \right.\nonumber\\
&+&\left.g_{\mathbf{k-k'}}\left[\psi_l^{A}(\mathbf{-k})\psi^A_l (\mathbf{k+q})+\psi_l^{B}(\mathbf{-k})\psi^B_l (\mathbf{k+q})\right] \left[\psi_{l'}^{A\dagger}(\mathbf{k'}) \psi_{l'}^{A\dagger}(\mathbf{-k'+q})+\psi_{l'}^{B\dagger}(\mathbf{k'}) \psi_{l'}^{B\dagger}(\mathbf{-k'+q})\right] + h.c.\right\}\nonumber
\end{eqnarray}

We expand the $g_\mathbf{k}$ form factors for small deviations away from the Dirac point $\mathbf{K}$; we set $F_0=g_\mathbf{2K}$, while the lowest order term in the expansion of $g_\mathbf{k-k'}$ away from the $\mathbf{K}$ point vanishes, introducing additional factors of momentum. This will introduce additional factors of temperature $T$ in the contribution to the thermal conductivity, and we therefore neglect the pair hopping term.

We work in the basis of the eigenstates of $H_0^l+H_{TRB}^l$. In the TR symmetric case, where $\Delta = m = 0$, the eigenstates are given by $a_\lambda(\mathbf{k}) = [\psi^{A}(\mathbf{k})+e^{i\phi_\mathbf{k}}\psi^{B}(\mathbf{k})]/\sqrt{2}$, where $\phi_\mathbf{k}$ is the angle between $k_x$ and $k_y$, and their energies are $\epsilon^\lambda_\mathbf{k} = \lambda v k$. For simplicity, in the analysis of the $Z_2$ QSL with a Fermi surface, we consider the regime $\Delta>m>0$, $m\gg\Delta-m$. In this limit, the eigenstates with energy close to the Fermi surface are located almost entirely on the $A$ sublattice, and we may ignore the sublattice degree of freedom; these eigenstaes are denoted by $a(\mathbf{k}) \sim \psi^{A}(\mathbf{k})$. In this basis, the single layer Green's functions are $G(\mathbf{k},i\nu_n) = [i\nu_n-k^2/2m^*+\mu]^{-1}$, 
with $\mu = \Delta-m$, and $m^* = v^2/m$.

\subsection{Generic inter-layer Hamiltonian}
\label{sec:generic}
The inter-layer coupling term (\ref{eq:inter}) is designed to maintain the exact solvability of the model, and for computational convenience. However, generically, we expect the largest components of the inter-layer coupling Hamiltonian to be quadratic in the spin operators. Within the Kitaev model, a quadratic inter-layer coupling term (such as a Heisenberg term, $J'_\perp \sum_{\alpha} S^\alpha_l S^\alpha_{l+1}$ with $\alpha = x$, $y$, $z$) does not merely create a pair of spinon excitations in each layer. Rather, it creates a pair of spinons and a pair of gapped vison (flux) excitations. In order to annihilate the pair of flux excitations and return to the low-energy subspace, we have to apply the $J'_\perp$ term again. Thus, it appears that the effective inter-plane interaction in the low-energy effective Hamiltonian~(\ref{couplings}) is proportional to $J_\perp \sim (J_\perp')^2/\Delta_v$, where $J_\perp'$ is the ``microscopic'' strength of the inter-layer coupling, and $\Delta_v$ is the vison gap. One may then worry that the effective inter-plane interaction $J_\perp$ is too small to contribute significantly to the inter-layer thermal conductivity.

However, we argue that for a generic intra-layer Hamiltonian that contains also non-Kitaev terms (such that even the intra-layer Hamiltonian is not exactly solvable), this is not the case; in fact, $J_\perp \propto J_\perp'$. This is because in the generic case, the vison excitations are not static, even within the single-layer Hamiltonian. A pair of vison excitations can annihilate each other without the need for another application of the inter-layer $J_\perp'$ term.

To illustrate this, consider the case where there is an additional intra-plane interaction:
\begin{equation}
H_{\Gamma} = \sum_{\langle i,j \rangle,l,\alpha,\beta} \Gamma_{\alpha\beta} S^\alpha_{i,l} S^\beta_{j,l},
\end{equation}
where $\langle i,j \rangle$ denotes two nearest-neighbor sites $i$, $j$ on the honeycomb lattice, and $\Gamma_{\alpha\beta}$ is a $3\times3 $ symmetric matrix. Such terms are present in real ``Kitaev materials''~\cite{KinHo2015}. 

Consider a quadratic inter-plane coupling term of the form $J_\perp' S^z_{1,l} S^z_{1,l+1}$, where the position of the site $1$ is indicated in Fig.~\ref{hexagon}. We can now derive the effective inter-plane coupling $J_\perp$ (that creates a pair of spinons in each layer and does not create any visons) perturbatively in both $J_\perp'$ and $\Gamma_{\alpha\beta}$. One can check explicitly that acting with the following sequence of operators:
\begin{equation}
S^z_{1,l} \big(S^x_{1,l}  S^x_{2,l} \big) \big(S^y_{1,l}  S^x_{2,l} \big) \times (l \rightarrow l+1),
\end{equation}
amounts, in a certain gauge, to acting $c_{1,l} c_{2,l} c_{1,l+1} c_{2,l+1}$ and not changing the number of visons in either layer. The strength of this term is $J_\perp \sim \frac{ J'_\perp (\Gamma_{xx} \Gamma_{xy})^2 }{\Delta_v^4}$. Thus, the term in the effective Hamiltonian that creates a pair of fermionic spinons in each of the adjacent layers $l$, $l+1$ is proportional to $J_\perp'$. Generically, there is no reason to expect $\Gamma_{\alpha\beta}$ to be much smaller in magnitude than $\Delta_v$, since both energy scales characterize the intra-plane Hamiltonian, and do not involve inter-layer coupling. We conclude that in a generic situation, $J_\perp$ and $J'_\perp$ are of the same order of magnitude.

\section{Evaluation of the integral in Eq.~(\ref{kappaz2clean})}
\label{app:integral}

The calculation proceeds similarly to the computation of the lifetime of a quasi-particle in a Fermi liquid. We integrate over $\mathbf{k}$, $\mathbf{k}'$ first, fixing $\mathbf{q}$.
Let us choose the axes such that $\mathbf{q}$ is in the $x$ direction.
For $q<2k_{F}$, there are pairs of points on the Fermi surface that
are connected by $\mathbf{q}$; we denote these points by $\mathbf{k}_{0}$,
$\mathbf{k}_{0}+\mathbf{q}$. Then, we parametrize

\begin{align}
\mathbf{k} & =\mathbf{k}_{0}+\delta\mathbf{k},\nonumber \\
\mathbf{k} & =-\mathbf{k}_{0}+\delta\mathbf{k}'.
\end{align}
It is convenient to linearize the dispersion around the Fermi surface;
then, to leading order in $\delta k$,

\begin{align}
\epsilon_{\mathbf{k}} & =v\delta k_{x}\sin\theta+v\delta k_{y}\cos\theta,\nonumber \\
\epsilon_{\mathbf{k}+\mathbf{q}} & =-v\delta k_{x}\sin\theta+v\delta k_{y}\cos\theta,\nonumber \\
\epsilon_{\mathbf{k}'} & =-v\delta k_{x}\sin\theta-v\delta k_{y}\cos\theta,\nonumber \\
\epsilon_{\mathbf{k}'-\mathbf{q}} & =v\delta k_{x}\sin\theta-v\delta k_{y}\cos\theta.
\end{align}
Here, $\sin\theta=\frac{q}{2k_{F}}$ (see Fig.~\ref{fig:FS}), and $v=\frac{k_{F}}{m^{*}}$
is the Fermi velocity. The integrals over $\delta\mathbf{k},$ $\delta\mathbf{k}'$
can now be performed easily, giving

\begin{align}
\kappa_{c} & =2\pi\frac{J_{\perp}^{2}}{T^{2}}\int d\epsilon_{1}d\epsilon_{2}d\epsilon_{3}(1-n_{F}(\epsilon_{1}))n_{F}(\epsilon_{2})(1-n_{F}(\epsilon_{3}))n_{F}(\epsilon_{1}-\epsilon_{2}+\epsilon_{3})\times(\epsilon_{1}-\epsilon_{2})^{2}\nonumber \\
 & \times\int\frac{d^{2}q}{(2\pi)^{6}}\frac{1}{v_{F}^{4}\cos^{2}\theta\sin^{2}\theta}.
\end{align}

The integral over $\epsilon_{1,2,3}$ can now be performed; by scaling,
this integral is proportional to $T^{5}$. Substituting $\theta$
with $q$, we get that

\begin{equation}
\kappa_{c}\propto\frac{J_{\perp}^{2}}{v_{F}^{4}}T^{3}\int_{0}^{2k_{F}}dq\frac{k_{F}^{2}}{q\left[1-q^{2}/\left(4k_{F}^{2}\right)\right]}.
\label{eq:kap}
\end{equation}
Using $\nu=\frac{k_{F}}{2\pi v_{F}}$, we arrive at Eq. (\ref{kappaz2clean}) in the text.
The logarithmic divergence comes from small $\mathbf{q}$ scattering,
as well as from scattering with momentum transfer close to $q=2k_F$. The inelastic lifetime of the
quasi-particles in each layer, $\tau\propto1/T^{2}$, is needed in
order to cut off this divergence. There is another contribution from
$q>2k_{F};$ this contribution is parametrically the same as Eq. (\ref{eq:kap}). 

\begin{figure}[h]
\centering{
\includegraphics[width=0.3\textwidth]{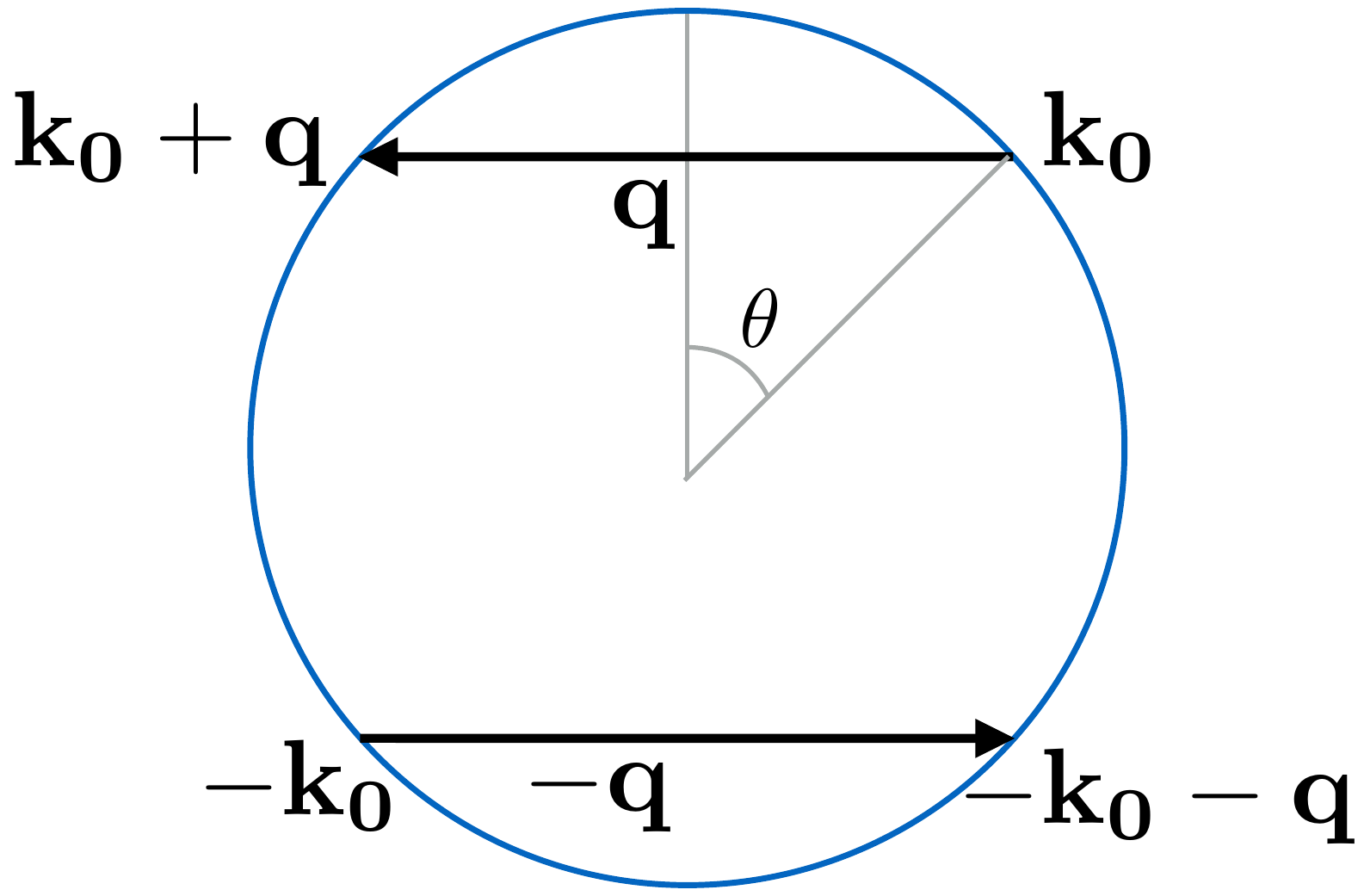}
\caption{Choice of $\mathbf{k}_0$ for a given $\mathbf{q}$ when integrating over $\mathbf{k}$, $\mathbf{k}'$ in Eq.~(\ref{kappaz2clean}). The blue circle is the Fermi surface.} 
\label{fig:FS}}
\end{figure}

\section{Disordered $Z_2$ QSL}
\subsection{Possible forms of disorder}\label{ap:disorder}
\label{sec:disorder_Z2}
In a TR invraiant system, we consider disorder in the spin-spin couplings $J$:
\begin{eqnarray}
H_{dis} = \sum_{<j,k>} \delta J_{jk} S_j^{\alpha_{jk}} S_k^{\alpha_{jk}},
\end{eqnarray}
where again $j,k$ are nearest neighbors and $\alpha_{jk}=x,y,z$, according to the type of link. In the Majorana fermion representation, this becomes
\begin{eqnarray}
H_{dis} = -\frac{1}{2}i\sum_{i} \delta J_{i,i+\delta} A_i B_{i+\delta} = v\int \frac{d^2kd^2k'}{(2\pi)^4}\psi^\dagger_{\mathbf{k}}\boldsymbol{\mathcal{A}_{\mathbf{k,k'}}\cdot \sigma} \psi_\mathbf{k'},
\end{eqnarray}
with $\mathbf{\mathcal{A}_{\mathbf{k,k'}}}^x = \frac{1}{J}\sum_{j,\delta}\Re[e^{-ik'\delta}\delta J_{j,\delta}e^{i(\mathbf{k-k'})R_j}], \mathbf{\mathcal{A}_{\mathbf{k,k'}}}^y = \frac{1}{J}\sum_{j,\delta}\Im[e^{-ik'\delta}\delta J_{j,\delta}e^{i(\mathbf{k-k'})R_j}]$; thus, the low energy, long-wavelength disorder is of the form of a random vector potential. 

%

Disorder which affects the next nearest neighbor hopping results from three-spin interaction terms in the original spin model; these terms break time reversal symmetry. Depending on the relative sign of the disorder between the $A$ and $B$ sublattices, just as in Eq.~(\ref{eq:nearestneighbors}), these terms will give a rise to a scalar potential term
\begin{equation}
H_{dis} = \int \frac{d^2kd^2k'}{(2\pi)^4}\mathcal{V}_{\mathbf{k,k'}}\psi^\dagger_{\mathbf{k}}\psi_\mathbf{k'},
\end{equation}
or a random mass term
\begin{equation}
H_{dis} = \int \frac{d^2kd^2k'}{(2\pi)^4}\psi^\dagger_{\mathbf{k}}\mathcal{M}_{\mathbf{k,k'}}\sigma^z \psi_\mathbf{k'}.
\end{equation}

\subsection{Inter-layer thermal conductivity}\label{scaling}
To compute the c-axis conductivity in a disordered, layered $Z_2$ QSL, we compute the rate at which pairs of spinon excitations tunnel between planes. 
This can done using the Fermi golden rule, analogously to Eq.~(\ref{kappaz2clean}), replacing the momentum eigenstates with eigenstates of the disordered intra-plane Hamiltonian.

The general expression for the thermal conductivity is
\begin{eqnarray}
\kappa_{c} & = & \frac{2\pi J_{\perp}^{2}}{ZT^{2}}\sum_{i,f}\sum_{l,l'}e^{-\frac{E_{i,l}}{T}-\frac{E_{i,l'}}{T}}(E_{i,l}-E_{f,l})^{2}
\vert\langle f\vert  H_\perp(l,l') \vert i\rangle\vert^{2}
 \times\delta(E_{i,l}+E_{i,l'}-E_{f,l}-E_{f,l}).
\end{eqnarray}
Here, $H_\perp(l,l')$ is the part of the inter-plane Hamiltonian that couples layers $l$ and $l'$. $|i,f\rangle$ are the initial and final many-body eigenstates of the system with $J_\perp=0$ (decoupled planes), with corresponding energies $E_{i,l}$ and $E_{f,l}$ (at layer $l$). Since we neglect intra-plane interactions, we can expand the fermionic spinon operators in the basis of the single-particle eigenstates of each layer (which include the effects of disorder):
\begin{equation}
\psi^\eta_l(x) = \sum_\lambda \varphi_{l,\lambda,\eta}(x)f_{\lambda,l},
\label{eq:eigen}
\end{equation}
where $\eta=A,B$ labels the sublattice, and $f_{\lambda,l}$ annihilates an eigenstate with energy $\xi_{\lambda,l}$, which has the wavefunction $\varphi_{l,\lambda,\eta}(x)$.

As in the main text, we will mostly work with an inter-layer Hamiltonian of the ``density-density'' form, $H_\perp(l,l') = \sum_{\eta,\eta'} \int d^2 x \, \psi^{\eta \dagger}_l(x) \psi^\eta_l(x) \psi^{\eta \dagger}_{l'}(x) \psi^\eta_{l'}(x)$, commenting along the way about other forms of $H_\perp$. 
Using Eq.~(\ref{eq:eigen}), we can write the disorder-averaged c-axis thermal conductivity as
\begin{eqnarray}
\kappa_{c} & = & \frac{2\pi J_{\perp}^{2}}{T^{2}}\sum_{\lambda_{1\dots  4}} \sum_{\eta_{1 \dots 4}} n_F(\xi_{l\lambda_1})\left[1-n_F(\xi_{l\lambda_3})\right]n_F(\xi_{l'\lambda_2})\left[1-n_F(\xi_{l'\lambda_4})\right](\xi_{l\lambda_1}-\xi_{l\lambda_3})^{2}\delta\left(\xi_{l\lambda_1}+\xi_{l'\lambda_2}-\xi_{l\lambda_3}-\xi_{l'\lambda_4}\right)\nonumber \\
 &  & \times\int d^{2}x\,d^{2}x' \left\langle\varphi_{l\lambda_1 \eta_1}^{*}(x)\varphi_{l\lambda_3 \eta_1}(x)\varphi_{l\lambda_1 \eta_2}(x')\varphi_{l\lambda_3 \eta_2}^{*}(x')\varphi_{l'\lambda_2 \eta_3}^{*}(x)\,\varphi_{l'\lambda_4\eta_3}(x)\varphi_{l'\lambda_2\eta_4}(x')\varphi_{l'\lambda_4\eta_4}^{*}(x') \right\rangle_{dis},
\label{eq:kappa_dis}
\end{eqnarray}
where $\langle \dots \rangle_{dis}$ represents disorder averaging.

We denote
\begin{equation}
g_{l}(q,\varepsilon,\varepsilon')=\sum_{\lambda_{1,2},\eta,\eta'}\int d^{2}x\,e^{-iq\cdot x}\left\langle \varphi_{l\lambda_1\eta}^{*}(x)\varphi_{l\lambda_2\eta}(x)\varphi_{l\lambda_1\eta'}(0)\varphi_{l\lambda_2\eta'}^{*}(0)\right\rangle _{dis}\delta(\varepsilon-\xi_{\l,\lambda_1})\delta(\varepsilon'-\xi_{l,\lambda_2}),
\end{equation}
such that the thermal conductivity is given by
\begin{eqnarray}\label{disorderedkappa}
\kappa_{c} &=& \frac{2\pi J_{\perp}^{2}}{T^{2}}\int d\epsilon_1d\epsilon_2d\omega n_F(\epsilon_1)\left[1-n_F(\epsilon_1+\omega)\right]n_F(\epsilon_2)\left[1-n_F(\epsilon_2-\omega)\right]\omega^{2}\nonumber\\
&\times& \int d^2q g_l(q, \epsilon_1,\epsilon_1+\omega) g_l(-q, \epsilon_2,\epsilon_2-\omega).
\end{eqnarray}

The remaining task is to compute the function $g_l(\mathbf{q},\varepsilon, \varepsilon')$ for a $Z_2$ QSL with either a Dirac spectrum or a Fermi surface.

%
\subsubsection{Disordered Dirac}\label{VPdisorder}
The properties of two-dimensional Dirac fermions coupled to a random vector potential, that corresponds to a disordered $Z_2$ Dirac QSL with time reversal symmetry, has been studied extensively in Ref.~\cite{Ludwig}. Here, we will briefly review some of the results of Ref.~\cite{Ludwig}, and use them to determine the scaling of the c-axis thermal conductivity with temperature.

Since the problem is non-interacting, the actions for different frequency modes decouple (before disorder averaging). The $\omega=0$ system is described by a fixed line of interacting theories in $d=1+1$ dimensions~\cite{Ludwig}. The frequency $\omega$ corresponds to a relevant operator with scaling dimension $2-z$, where $z=1+\Delta_{A}/\pi$ is the dynamical critical exponent, and $\Delta_{A}$ is the disorder strength, defined in Eq.~(\ref{eq:Delta_A}). I.e., under scaling, $q\rightarrow q' = q/b$, $\omega\rightarrow\omega' = \omega/b^{z}$. The fixed line is characterized by $\omega/T$ scaling. 

The function $g_{l}(q,\varepsilon,\varepsilon')$ satisfies the scaling relation  


\begin{equation}
g_{l}(q,\varepsilon_{1},\varepsilon_{2})=b^{-y} g_l\left(bq, b^z \varepsilon_{1},b^z \varepsilon_{2}\right),\label{eq:scaling-g}
\end{equation}
where $y$ is a critical exponent related to the scaling dimension of the fermion density operator, which we compute below,
and $b$ is a rescaling factor. Choosing $b = |\varepsilon|^{-1/z}$, we get that $g_l$ can be written as
\begin{equation}
g_l(q,\varepsilon_1,\varepsilon_2) = |\varepsilon_1|^{y/z} \Phi\left(\frac{q}{|\varepsilon_1|^{1/z}}, \frac{\varepsilon_2}{|\varepsilon_1|}\right),
\label{eq:g_scale}
\end{equation}
where $\Phi$ is a universal scaling function.

To determine $y$, we notice that $g_{l}$ is related to the density-density correlator:
\begin{eqnarray}\label{eq:nq}
\chi(q,\omega_n) & \equiv&\langle n(q,\omega_n)n(-q,\omega_n)\rangle \nonumber \\
&= & \int d^{2}x\,e^{-iq\cdot x}\sum_{\eta,\eta'}\sum_{\alpha,\gamma}\frac{n_F(\varepsilon_{\alpha})-n_F(\varepsilon_{\gamma})}{i\omega_n -\varepsilon_{\alpha}+\varepsilon_{\gamma}}\left\langle \varphi_{l\alpha\eta}^{*}(x)\varphi_{l\gamma\eta}(x)\varphi_{l\alpha\eta'}(0)\varphi_{l\gamma\eta'}^{*}(0)\right\rangle _{dis}\nonumber \\
&= & \int d\varepsilon d\varepsilon'\frac{n_F(\varepsilon)-n_F(\varepsilon')}{i\omega_n-\varepsilon+\varepsilon'}g_{l}(q,\varepsilon,\varepsilon'), 
\label{eq:Pi}
\end{eqnarray}
where $n_F(\varepsilon)$ is the Fermi function. Using Eq.~(\ref{eq:g_scale}), this can be written as
\begin{equation}
\chi(q,\omega_n)=  T^{1+y/z} \int d\xi d\xi'\frac{n_F(T\xi)-n_F(T\xi')}{i\omega_n-\xi+\xi'} |\xi|^{y/z} \Phi\left(\frac{q}{T|\xi|^{1/z}},\frac{\xi'}{|\xi|}\right). 
\end{equation}
Here, we have used Eq.~(\ref{eq:g_scale}) and performed a change of variables, $\varepsilon = T \xi$, $\varepsilon' = T \xi'$.
On the other hand, $\chi(q,\omega_n=0)$ can be expressed as
\begin{eqnarray}
\chi(q,\omega_n  =0,T) & = & \int_{0}^{\beta}d\tau\int d^{2}x\, e^{-iq\cdot x}\langle n_l (x,\tau)n_l(0,0)\rangle \nonumber \\ 
& = & T\sum_{\nu,\nu'}\int d^{2}x\, e^{-iq\cdot x}
\langle n_{l,\nu}(x) n_{l,\nu'}(0)\rangle \nonumber\\ 
& = & T\sum_{\nu,\nu'}\int d^{2}x\, e^{-iq\cdot x}
 b^{-2(2-z)} \langle n_{l,\nu}(x/b) n_{l,\nu'}(0)\rangle \nonumber \\ 
& = & b^{2(z - 1)}T\,\chi(bq,\omega=0,b^{z}T).
\label{eq:Pi_scale}
\end{eqnarray}
Here, $n_{l,\nu}(x)$ is the Matsubara frequency $\nu$ component of the density operator in layer $l$~\footnote{The factor of $T$ in front of $\chi$ in Eq.~(\ref{eq:Pi_scale}) comes our convention of the Matsubara fermionic fields: $\psi(i\nu_n) = T^{1/2} \int_0^\beta d\tau e^{i\nu_n \tau} \psi(\tau)$. In this convention, there are no factors of $T$ in the quadratic part of the action.}. In the second to last line we have applied scaling to the correlation function $\langle n_{l,\nu}(x) n_{l,\nu'}(0)\rangle$, using the fact that the scaling dimension of $n_{l,\nu}(x)$ is $2-z$.\cite{Ludwig} 

Choosing $b=T^{-1/z}$ in Eq.~(\ref{eq:Pi_scale}), we get that $\chi(q,\omega_n  =0,T) = T^{\frac{2-z}{z}} \Psi(q/T^{1/z})$, where $\Psi$ is a scaling function. 
Comparing this to Eq.~(\ref{eq:Pi}), we can extract the exponent $y$:
\begin{equation}
y = 2(1-z).
\end{equation}

Now, we are in a position to find the scaling of the c-axis thermal conductivity with temperature. Inserting Eq.~(\ref{eq:g_scale}) into Eq.~(\ref{disorderedkappa}) results in
\begin{eqnarray}
\kappa_{c} & = & \frac{2\pi J_{\perp}^{2}}{T^{2}} \int d\varepsilon_{1}d\varepsilon_{2}d\omega\,n(\varepsilon_{1})\left[1-n(\varepsilon_{1}+\omega)\right]n(\varepsilon_{2})\left[1-n(\varepsilon_{2}-\omega)\right]\omega^{2}\nonumber \\
 &  & \times\int d^{2}q\,
 |\varepsilon_1|^{y/z}\Phi\left(\frac{q}{|\varepsilon_1|^{1/z}},\frac{\varepsilon_{1}-\omega}{|\varepsilon_1|}\right)
 |\varepsilon_2|^{y/z}\Phi\left(-\frac{q}{|\varepsilon_2|^{1/z}},\frac{\varepsilon_{2}-\omega}{|\varepsilon_2|}\right)
\end{eqnarray}

 Rescaling the integral,  $\tilde{\omega}=\omega/T$, $\tilde{\varepsilon}_{1,2}=\varepsilon_{1,2}/T$, and $\tilde{q}=q/T^{1/z}$, we get that
\begin{equation}
\kappa_{c}\sim\frac{2\pi J_{\perp}^{2}}{T^{2}}T^{2/z}\left(T^{3}\right)\left(T^{2}\right)\left(T^{4(1-z)/z}\right)\sim J_{\perp}^{2}T^{(6-z)/z}.
\end{equation}
This result coincides with that of the clean case in the limit $\Delta_{A}\rightarrow0$ (i.e. $z=1+\Delta_{A}/\pi\rightarrow1$).

The analysis above has been done for an inter-plane interaction of the density-density form. A similar analysis can be done for any quartic inter-plane interaction. The only difference is the scaling dimension of the fermion bilinear operator that appears in the interaction term, that can be determined using the methods of Ref.~\cite{Ludwig}. It turns out, however, that for any $\Delta_A>0$, the density operator is the fermion bilinear with the smallest scaling dimension. Hence a density-density interaction gives the dominant contribution to $\kappa_c$ at low temperatures.

\subsubsection{Disordered FS}
\label{sec:disordered_FS}
In the disordered FS case, we know that the density-density correlation function takes a diffusive form at small $q,\omega_n$:
\begin{equation}
\chi(q,\omega_n)  = \nu\frac{ Dq^2}{|\omega_n|+Dq^2}.
\label{eq:diffuse}
\end{equation}
where $D$ is the diffusion constant, and $\nu$ is the density of states at the Fermi level. 
Comparing this to Eq.~(\ref{eq:Pi}), we deduce that $g_l(q,\varepsilon,\varepsilon')$ should satisfy the following scaling relation:
\begin{equation}
g_l(q,\varepsilon,\varepsilon') = b^2 g_l(bq, b^2\varepsilon, b^2 \varepsilon').
\end{equation}
Hence, $g_l(q,\varepsilon,\varepsilon')$ can be written as
\begin{equation}
g_l(q,\varepsilon,\varepsilon') = \frac{\nu}{Dq^2} \Omega\left(\frac{\varepsilon}{Dq^2}, \frac{\varepsilon'}{Dq^2} \right),
\end{equation}
where $\Omega(\xi,\xi')$ is a dimensionless scaling function. 

We may now use this form in Eq.~(\ref{disorderedkappa}) to get:
\begin{eqnarray}
\kappa_{c} & = & \frac{2\pi J_{\perp}^{2}}{T^{2}} \int d\varepsilon_{1}d\varepsilon_{2}d\omega\,n(\varepsilon_{1})\left[1-n(\varepsilon_{1}+\omega)\right]n(\varepsilon_{2})\left[1-n(\varepsilon_{2}-\omega)\right]\omega^{2}\nonumber \\
 &  & \times\int d^{2}q\,
\frac{\nu^2}{D^2q^4}\Omega\left(\frac{\varepsilon_1}{Dq^2},\frac{\varepsilon_{1}-\omega}{Dq^2}\right)
\Omega\left(\frac{\varepsilon_2}{Dq^2},\frac{\varepsilon_{2}-\omega}{Dq^2}\right).
\end{eqnarray}
Changing variables to $\tilde{\varepsilon}_{1,2} = \varepsilon_{1,2}/T$, $\tilde{q} = q/\sqrt{DT}$, we obtain
\begin{equation}
\kappa_c \sim J_\perp^2 \nu^2 \frac{T^2}{D}.
\end{equation}
This result coincides with the result of the Kubo formula calculation described in the main text.

%

\section{Kubo formula for the thermal conductivity}\label{ap:Matsubara}

\subsection{Thermal current operator}\label{ap:current}
The systems we consider consist of layers of quasi-2D QSLs, described by the in-plane Hamiltonian $H^l$, and coupled by interplane hopping terms which may be written as sums of terms of the form $H_\perp=J_\perp O^l O^{l+1}$, where $O^l$ is composed of operators of the $l$ level only. The energy density of a single layer $l$ is thus (for a single term $H_\perp$; the extension to a sum of terms is straightforward) 
\begin{eqnarray}
E^l =H^l+\frac{1}{2}J_\perp\left\{O^lO^{l-1}+O^lO^{l+1}\right\}.
\end{eqnarray}
Its time derivative is then
\begin{eqnarray}
\dot{E}^l &=& iJ_\perp\left\{[O^l,H^l]O^{l-1}+[O^l,H^l]O^{l+1}+\frac{1}{2}[H^l,O^l]O^{l-1}+\frac{1}{2}[H^l,O^l]O^{l+1}+\frac{1}{2}[H^{l-1},O^{l-1}]O^{l}+\frac{1}{2}[H^{l+1},O^{l+1}]O^{l}\right\}\nonumber\\
&+&O(J_\perp^2)\\
&=&\frac{1}{2}iJ_\perp\left\{[H^{l-1},O^{l-1}]O^l-[H^l,O^l]O^{l-1}+[H^{l+1},O^{l+1}]O^l-[H^{l},O^{l}]O^{l+1}\right\}+O(J_\perp^2)\\
&=&\frac{1}{2}J_\perp\left\{\dot{O}^{l-1}O^l-\dot{O}^lO^{l-1}+\dot{O}^{l+1}O^l-\dot{O}^{l}O^{l+1}\right\}+O(J_\perp^2)
\end{eqnarray}
where in the last equality we have used the fact that to lowest order in $J_\perp$, $\dot{O}^{l}=i[H^l,O^l]$.

The thermal current operator is given by \cite{Mahan}
\begin{eqnarray}
J^Q = \sum_l l \dot{E}^l,
\end{eqnarray}
and therefore
\begin{eqnarray}\label{JQ}
J^Q &=& \frac{1}{2}J_\perp\sum_l\left\{(l+1)\dot{O}^{l}O^{l+1}-l\dot{O}^lO^{l-1}+(l-1)\dot{O}^{l}O^{l-1}-l\dot{O}^{l}O^{l+1}\right\}+O(J_\perp^2)\nonumber\\
&=&\sum_l J^Q_{l,l+1}+O(J_\perp^2)
\end{eqnarray}

with $J^Q_{l,l+1} = \frac{1}{2}J_\perp\left(\dot{O}^{l}O^{l+1}-O^l\dot{O}^{l+1}\right)$. This formula satisfies the continuity equation $J^Q_{l,l+1}-J^Q_{l-1,l} = \dot{E}_l$.

\subsection{Inter-layer thermal current for layered Kitaev honeycomb model}
\label{sec:inter_current}
For the $Z_2$ system, the coupling term is given by Eq.~(\ref{hperpz2})
\begin{eqnarray}
H_\perp &=& \frac{1}{2}{J_\perp}F_0\sum_{l,l' = l\pm1}
\int \frac{d^2k}{(2\pi)^2}\frac{d^2k'}{(2\pi)^2}\frac{d^2q}{(2\pi)^2}\\
&\times&\left[ \psi^{A\dagger}_l(\mathbf{k})
\psi^A_{l}({\mathbf{k}+\mathbf{q}})\psi^{A\dagger}_{l'}(\mathbf{k'})
\psi^{A}_{l'}({\mathbf{k'}-\mathbf{q}})  +  (A^l\rightarrow B^l) + (A^{l'}\rightarrow B^{l'}) + (A^l\rightarrow B^l \& A^{l'}\rightarrow B^{l'})\right]\nonumber
\end{eqnarray}

and therefore
\begin{eqnarray}
J^Q &=& \frac{1}{2}{J_\perp}F_0\int  \frac{d^2k}{(2\pi)^2}\frac{d^2k'}{(2\pi)^2}\frac{d^2q}{(2\pi)^2} \sum_l \partial_t\left[\psi^{A\dagger}_l(\mathbf{k})
\psi^{A}_l({\mathbf{k}+\mathbf{q}})\right]\\
&\times&\left[\left(\psi^{A\dagger}_{l+1}(\mathbf{k'})
\psi^A_{l+1}({\mathbf{k'}-\mathbf{q}})-\psi^{A\dagger}_{l-1}(\mathbf{k'})
\psi^{A}_{l-1}({\mathbf{k'}-\mathbf{q}})\right)  +(A^l\rightarrow B^l) + (A^{l'}\rightarrow B^{l'}) + (A^l\rightarrow B^l \, \& \, A^{l'}\rightarrow B^{l'})\right]\nonumber
\end{eqnarray}
Fourier transforming with respect to imaginary time results in
\begin{eqnarray}
J^Q(i\omega_n) &=& \frac{1}{2}{J_\perp}F_0 \int  \frac{d^2k}{(2\pi)^2}\frac{d^2k'}{(2\pi)^2}\frac{d^2q}{(2\pi)^2} \frac{1}{\beta^3}\sum_{\nu_n, \nu_m, \Omega_n} \Omega_n \sum_l \left[\psi^{A\dagger}_l(\mathbf{k},i\nu_n)
\psi^A_{l}({\mathbf{k}+\mathbf{q}},i\nu_n+i\Omega_n)\right]\times\nonumber\\
&&\left[\left(\psi^{A\dagger}_{l+1}(\mathbf{k'},i\nu_m)
\psi^A_{l+1}({\mathbf{k'}-\mathbf{q}},i\nu_m-i\Omega_n+i\omega_n)-(l+1)\rightarrow(l-1)\right)\right.\nonumber\\
&+&\left. (A^l\rightarrow B^l) + (A^{l'}\rightarrow B^{l'}) + (A^l\rightarrow B^l \& A^{l'}\rightarrow B^{l'})\right]
\end{eqnarray}
We again revert to the complex fermion representation, and consider states only near the Dirac point $\mathbf{K}$. Transforming to the eigenstates of $H$ yields, for $m = 0$, 
\begin{eqnarray}
J^Q (i\omega_n) 
& = & \frac{1}{2^5}{J_\perp}\sum_{l,\eta=\pm 1}\eta\int \frac{d^2k}{(2\pi)^2}\frac{d^2k'}{(2\pi)^2}\frac{d^2q}{(2\pi)^2} \sum_{\lambda_1...\lambda_4=\pm1}\frac{1}{{\beta^3}}\sum_{\nu_n,\nu_m,\Omega_n}\Omega_n\\
& \times&{F}_{\mathbf{k},\mathbf{k'},\mathbf{q}}^{\lambda_1...\lambda_4}
a_{\lambda_1}^{l\dagger}(\mathbf{k},i\nu_n)
a_{\lambda_2}^{l}(\mathbf{k}+\mathbf{q},i\nu_n+i\Omega_n+i\omega_n)a_{\lambda_3}^{l+\eta\dagger}(\mathbf{k'},i\nu_m)
a_{\lambda_4}^{l+\eta}(\mathbf{k'}-\mathbf{q},i\nu_m-i\Omega_n)\nonumber,
\end{eqnarray}
where ${F}_{\mathbf{k},\mathbf{k'},\mathbf{q}}^{\lambda_1...\lambda_4}$ corresponds to the transformation of $F_0$ from the sublattice to the eigenstate basis,
while for the case $\Delta>m>0, m\gg\Delta-m$, we neglect the contribution of the the $B$ sublattice and get a simpler expression
\begin{eqnarray}
J^Q (i\omega_n) 
& = & \frac{1}{2^5}{J_\perp}F_0\sum_{l,\eta=\pm 1}\eta\int \frac{d^2k}{(2\pi)^2}\frac{d^2k'}{(2\pi)^2}\frac{d^2q}{(2\pi)^2} \frac{1}{{\beta^3}}\sum_{\nu_n,\nu_m,\Omega_n}\Omega_n\\
& \times&
a^{l\dagger}(\mathbf{k},i\nu_n)
a^{l}(\mathbf{k}+\mathbf{q},i\nu_n+i\Omega_n+i\omega_n)a^{l+\eta\dagger}(\mathbf{k'},i\nu_m)
a^{l+\eta}(\mathbf{k'}-\mathbf{q},i\nu_m-i\Omega_n).\nonumber
\end{eqnarray}

\subsection{Thermal conductivity}

The thermal conductivity is given by \cite{Luttinger1, Luttinger2, Shastry}
\begin{eqnarray}
\kappa = \frac{-1}{T}\lim_{\omega\to 0} \frac{\Im\left[\Pi(\omega)\right]}{\omega},
\end{eqnarray}
where $\Pi(\omega)$ is the retarded thermal-current thermal-current correlation function. Using Eq.~(\ref{currentoperator}), and an extended definition of the four point correlation function in Eq.~(\ref{upsilon}) (in the TR broken case, there is only a single band $\lambda$ that crosses the Fermi surface)
\begin{equation}\label{vertex}
\begin{split}
&\Upsilon_{\lambda_1,\lambda_2}(\mathbf{k}_1,\mathbf{k}_1',\mathbf{q};i\nu_n,i\nu_m)\\
& =\left\langle a^\dagger_{\lambda_1}(\mathbf{k}_1,i\nu_n)a_{\lambda_2}(\mathbf{k}_1+\mathbf{q},i\nu_m)a^\dagger_{\lambda_2}(\mathbf{k}'_1+\mathbf{q},i\nu_m)a_{\lambda_1}(\mathbf{k}_1',i\nu_n)\right\rangle,
\end{split}
\end{equation}
we get

\begin{eqnarray}\label{kuboz2}
\Pi(i\omega_n) &=& \frac{1}{64}J_\perp^2\frac{1}{\beta}\sum_{\Omega_n}\Omega_n^2\int \frac{d^2q}{(2\pi)^2}\frac{1}{\beta}\sum_{\nu_n}\int \frac{d^2k_1d^2k_2}{(2\pi)^4}\frac{1}{\beta}\sum_{\nu_m}\int \frac{d^2k'_1d^2k'_2}{(2\pi)^4}\\
&\times&\sum_{\lambda_1...\lambda_4}F^{\lambda_1,\lambda_2,\lambda_3,\lambda_4}_{\mathbf{k_1,k_1',q}}F^{\lambda_2,\lambda_1,\lambda_4,\lambda_3}_{\mathbf{k_2,k_2',-q}}\Upsilon_{\lambda_2,\lambda_1}(\mathbf{k_1,k_2,q};-i\nu_n-i\Omega_n-i\omega_n,-i\nu_n)
\Upsilon_{\lambda_3,\lambda_4}(\mathbf{k'_1,k'_2,-q};i\nu_m,i\nu_m-i\Omega_n)\nonumber
\end{eqnarray}

The function $\Upsilon_\mathbf{k}(z,z+i\Omega_n)$ (supressing the momentum and $\lambda$ dependence for clarity) has branch cuts for $\Im[z]=0, \Im[z]=-i\Omega_n$; using the usual contour integration method, as explained in \cite{Mahan}, for example, it can be shown that 

\begin{eqnarray}
&&\frac{1}{\beta}\sum_{\nu_n}\Upsilon(z,z+i\Omega_n) = \\
&&\int\frac{d\epsilon_1}{2\pi i} n_F(\epsilon_1)\left[\Upsilon_\mathbf{k}(\epsilon_1^+, \epsilon_1+i\Omega_n)-\Upsilon_\mathbf{k}(\epsilon_1^-, \epsilon_1+i\Omega_n)\right]+\int\frac{d\epsilon_1}{2\pi i} n_F(\epsilon_1)\left[\Upsilon_\mathbf{k}( \epsilon_1-i\Omega_n, \epsilon_1^+)-\Upsilon_\mathbf{k}(\epsilon_1-i\Omega_n, \epsilon_1^-)\right]\nonumber
\end{eqnarray}

where $\epsilon^\pm = \epsilon\pm i\delta$, with $\delta$ a positive infinitesimal. Performing first the summations over $\nu_n$ and $\nu_m$, which result in integrations over $\epsilon_1, \epsilon_3$, the summation over the bosonic frequencies $\Omega_n$ gives (again integrating along the branch cuts of the $\Upsilon$ functions)
	
Performing first the summations over $\nu_n$ and $\nu_m$, which result in integrations over $\epsilon_1, \epsilon_3$, the summation over the bosonic frequencies $\Omega_n$ gives (again integrating along the brach cuts of the $\Upsilon$ functions)
\begin{eqnarray}
&&Q(i\omega_n) \equiv \frac{1}{\beta^3}\sum_{\nu_n, \nu_m, \Omega_n} \Omega_n^2 \Upsilon_{\mathbf{k}}(i\nu_n, i\nu_n+i\Omega_n) \Upsilon_{\mathbf{k'}}(i\nu_m, i\nu_m-i\Omega_n+i\omega_n) = \\
&& \int\frac{d\epsilon_1}{2\pi i}\int\frac{d\epsilon_3}{2\pi i}n_F(\epsilon_1)n_F(\epsilon_3)\int\frac{d\epsilon_2}{2\pi i}\times\nonumber\\
&&n_B(x)x^2\left[\Upsilon_{\mathbf{k}}(\epsilon_1^+, \epsilon_2^+)-\Upsilon_{\mathbf{k}}(\epsilon_1^+, \epsilon_2^-)-\Upsilon_{\mathbf{k}}(\epsilon_1^-, \epsilon_2^+)+\Upsilon_{\mathbf{k}}(\epsilon_1^-, \epsilon_2^-)\right]\times\nonumber\\
&&\left[\Upsilon_{\mathbf{k'}}(\epsilon_3^+, -x+\epsilon_3+i\omega_n)-\Upsilon_{\mathbf{k'}}(\epsilon_3^-, -x+\epsilon_3+i\omega_n)+\Upsilon_{\mathbf{k'}}(x+\epsilon_3-i\omega_n, \epsilon_3^+)-\Upsilon_{\mathbf{k'}}(x+\epsilon_3-i\omega_n, \epsilon_3^-)\right]\nonumber\\
&&-n_B(-x)x^2 \left[\Upsilon_{\mathbf{k}}(\epsilon_2^+, \epsilon_1^+)-\Upsilon_{\mathbf{k}}(\epsilon_2^+, \epsilon_1^-)-\Upsilon_{\mathbf{k}}(\epsilon_2^-, \epsilon_1^+)+\Upsilon_{\mathbf{k}}(\epsilon_2^-, \epsilon_1^-)\right]\times\nonumber\\
&&\left[\Upsilon_{\mathbf{k'}}(\epsilon_3^+, x+\epsilon_3+i\omega_n)-\Upsilon_{\mathbf{k'}}(\epsilon_3^-, x+\epsilon_3+i\omega_n)+\Upsilon_{\mathbf{k'}}(-x+\epsilon_3-i\omega_n, \epsilon_3^+)-\Upsilon_{\mathbf{k'}}(-x+\epsilon_3-i\omega_n, \epsilon_3^-)\right]\nonumber\\
&&+n_B(y)(y+i\omega_n)^2 \left[\Upsilon_{\mathbf{k'}}(\epsilon_2^+, \epsilon_3^+)-\Upsilon_{\mathbf{k'}}(\epsilon_2^+, \epsilon_3^-)-\Upsilon_{\mathbf{k'}}(\epsilon_2^-, \epsilon_3^+)+\Upsilon_{\mathbf{k'}}(\epsilon_2^-, \epsilon_3^-)\right]\times\nonumber\\
&&\left[\Upsilon_{\mathbf{k}}(\epsilon_1^+, \epsilon_1+y+i\omega_n)-\Upsilon_{\mathbf{k}}(\epsilon_1^-, \epsilon_1+y+i\omega_n)+\Upsilon_{\mathbf{k}}(\epsilon_1-y-i\omega_n, \epsilon_1^+)-\Upsilon_{\mathbf{k}}(\epsilon_1-y-i\omega_n, \epsilon_1^-)\right]\nonumber\\
&&-n_B(-y)(-y+i\omega_n)^2\left[\Upsilon_{\mathbf{k'}}(\epsilon_3^+, \epsilon_2^+)-\Upsilon_{\mathbf{k'}}(\epsilon_3^+, \epsilon_2^-)-\Upsilon_{\mathbf{k'}}(\epsilon_3^-, \epsilon_2^+)+\Upsilon_{\mathbf{k'}}(\epsilon_3^-, \epsilon_2^-)\right]\times\nonumber\\
&&\left[\Upsilon_{\mathbf{k}}(\epsilon_1^+, \epsilon_1-y+i\omega_n)-\Upsilon_{\mathbf{k}}(\epsilon_1^-, \epsilon_1-y+i\omega_n)+\Upsilon_{\mathbf{k}}(\epsilon_1+y-i\omega_n, \epsilon_1^+)-\Upsilon_{\mathbf{k}}(\epsilon_1+y-i\omega_n, \epsilon_1^-)\right],\nonumber
\end{eqnarray}
with $x = \epsilon_2-\epsilon_1, y = \epsilon_2-\epsilon_3$.

Performing the analytical continuation by replacing $i\omega_n\rightarrow \omega+i\delta$ and massaging the expressions a bit results in
\begin{eqnarray}
&&Q(\omega)= \int\frac{d\epsilon_1}{2\pi i}\int\frac{d\epsilon_3}{2\pi i}\int\frac{d\epsilon_2}{2\pi i}\times\nonumber\\
&&n_B(x)x^2\left[n_F(\epsilon_1)-n_F(\epsilon_2)\right]\left[\Upsilon_{\mathbf{k}}(\epsilon_1^+, \epsilon_2^+)-\Upsilon_{\mathbf{k}}(\epsilon_1^+, \epsilon_2^-)+c.c.\right]\times\nonumber\\
&&\left[n_F(\epsilon_3)\left(\Upsilon_{\mathbf{k'}}(\epsilon_3^+, \epsilon_3^+-x+\omega)-\Upsilon_{\mathbf{k'}}(\epsilon_3^-, \epsilon_3^+-x+\omega)\right)+n_F(\epsilon_3-x+\omega)\left(\Upsilon_{\mathbf{k'}}(\epsilon_3^-, \epsilon_3^+-x+\omega)-\Upsilon_{\mathbf{k'}}(\epsilon_3^-, \epsilon_3^--x+\omega)\right)\right]\nonumber\\
&&-n_B(y)(-y+\omega)^2\left[n_F(\epsilon_3)-n_F(\epsilon_2)\right]\left[\Upsilon_{\mathbf{k'}}(\epsilon_3^+, \epsilon_2^+)-\Upsilon_{\mathbf{k'}}(\epsilon_3^-, \epsilon_2^+)-c.c\right]\times\nonumber\\
&&\left[n_F(\epsilon_1)\left(\Upsilon_{\mathbf{k}}(\epsilon_1^+, \epsilon_1^+-y+\omega)-\Upsilon_{\mathbf{k}}(\epsilon_1^-, \epsilon_1-y+\omega)\right)+n_F(\epsilon_1-y+\omega)\left(\Upsilon_{\mathbf{k}}(\epsilon_1^-, \epsilon_1^+-y+\omega)-\Upsilon_{\mathbf{k}}(\epsilon_1^-, \epsilon_1^--y+\omega)\right)\right]\nonumber
\end{eqnarray}

The imaginary part of the above expression is (using the fact that $\Upsilon_{\mathbf{k}}(\epsilon_1^+, \epsilon_2^+) = \Upsilon_{\mathbf{k}}(\epsilon_1^-, \epsilon_2^-)^*$ and
$\Upsilon_{\mathbf{k}}(\epsilon_1^+, \epsilon_2^-) =  \Upsilon_{\mathbf{k}}(\epsilon_1^-, \epsilon_2^+)^*$)
\begin{eqnarray}
&&\Im[Q(\omega)]= \int\frac{d\epsilon_1}{2\pi i}\int\frac{d\epsilon_3}{2\pi i}\int\frac{d\epsilon_2}{2\pi i}\times\nonumber\\
&&2\Re\left[\Upsilon_{\mathbf{k}}(\epsilon_1^+, \epsilon_2^+)-\Upsilon_{\mathbf{k}}(\epsilon_1^+, \epsilon_2^-)\right]\times\Re\left[\Upsilon_{\mathbf{k'}}(\epsilon_3^+, \epsilon_3^+-x+\omega)-\Upsilon_{\mathbf{k'}}(\epsilon_3^-, \epsilon_3^+-x+\omega)\right]\times\nonumber\\
&&\left[n_F(\epsilon_3)-n_F(\epsilon_3-x+\omega)\right]\times\left[n_F(\epsilon_1)-n_F(\epsilon_2)\right]n_B(x)x^2\nonumber\\
&&-2\Re\left[\Upsilon_{\mathbf{k'}}(\epsilon_3^+, \epsilon_2^+)-\Upsilon_{\mathbf{k'}}(\epsilon_3^-, \epsilon_2^+)\right]\times\Re\left[\Upsilon_{\mathbf{k}}(\epsilon_1^+, \epsilon_1^+-y+\omega)-\Upsilon_{\mathbf{k}}(\epsilon_1^-, \epsilon_1^+-y+\omega)\right]\times\nonumber\\
&&\left[n_F(\epsilon_1)-n_F(\epsilon_1-y+\omega)\right]\left[n_F(\epsilon_1)-n_F(\epsilon_2)\right]n_B(-y)(-y+\omega)^2\nonumber
\end{eqnarray}
In the second line replace $\epsilon_2\rightarrow\epsilon_1-\epsilon_2+\epsilon_3+\omega$ to get
\begin{eqnarray}
&&\Im[Q(\omega)]= \int\frac{d\epsilon_1}{2\pi i}\int\frac{d\epsilon_3}{2\pi i}\int\frac{d\epsilon_2}{2\pi i}\times\nonumber\\
&&2\Re\left[\Upsilon_{\mathbf{k}}(\epsilon_1^+, \epsilon_2^+)-\Upsilon_{\mathbf{k}}(\epsilon_1^+, \epsilon_2^-)\right]\times\Re\left[\Upsilon_{\mathbf{k'}}(\epsilon_3^+, \epsilon_1-\epsilon_2+\epsilon_3^++\omega)-\Upsilon_{\mathbf{k'}}(\epsilon_3^-, \epsilon_1-\epsilon_2+\epsilon_3^++\omega)\right]\times\nonumber\\
&&(\epsilon_2-\epsilon_1)^2\left[n_F(\epsilon_1)-n_F(\epsilon_2)\right]\times\left[n_F(\epsilon_3)-n_F(\epsilon_3-\epsilon_2+\epsilon_1+\omega)\right]\left[n_B(\epsilon_2-\epsilon_1)-n_B(\epsilon_2-\epsilon_1-\omega)\right]
\end{eqnarray}
Therefore, using the identity
\begin{eqnarray}
\left[n_F(\epsilon_1)-n_F(\epsilon_2)\right]\left[n_F(\epsilon_3)-n_F(\epsilon_3-\epsilon_2+\epsilon_1)\right]\frac{\partial}{\partial\epsilon}n_B(\epsilon_2-\epsilon_1) = \frac{1}{T}(1-n_F(\epsilon_1))n_F(\epsilon_2)(1-n_F(\epsilon_3))n_F(\epsilon_1-\epsilon_2+\epsilon_3),\nonumber\\
\end{eqnarray}
we get
\begin{eqnarray}
&&\lim_{\omega\to 0}\frac{\Im[Q(\omega)]}{\omega}=\frac{1}{T}\int\frac{d\epsilon_1}{2\pi i}\int\frac{d\epsilon_3}{2\pi i}\int\frac{d\epsilon_2}{2\pi i}\times\nonumber\\
&&2\Re\left[\Upsilon_{\mathbf{k}}(\epsilon_1^+, \epsilon_2^+)-\Upsilon_{\mathbf{k}}(\epsilon_1^+, \epsilon_2^-)\right]\times\Re\left[\Upsilon_{\mathbf{k'}}(\epsilon_3^+, \epsilon_1-\epsilon_2+\epsilon_3^++\omega)-\Upsilon_{\mathbf{k'}}(\epsilon_3^-, \epsilon_1-\epsilon_2+\epsilon_3^++\omega)\right]\times\nonumber\\
&&(\epsilon_2-\epsilon_1)^2\times(1-n_F(\epsilon_1))n_F(\epsilon_2)(1-n_F(\epsilon_3))n_F(\epsilon_1-\epsilon_2+\epsilon_3).
\end{eqnarray}
Inserting this into the formula for $\kappa$, Eq.~(\ref{kubo}), and writing the momentum and band dependence explicitly, results in 

\begin{eqnarray}\label{kappa}
\kappa &\sim& 
  \frac{J_\perp^2}{ T^2}\int \frac{d^2k_1}{(2\pi)^2}\frac{d^2k_2}{(2\pi)^2}\frac{d^2k_1'}{(2\pi)^2}\frac{d^2k_2'}{(2\pi)^2}\frac{d^2q}{(2\pi)^2} 
\int\frac{d\epsilon_1}{2\pi}\int\frac{d\epsilon_2}{2\pi}\int\frac{d\epsilon_3}{2\pi}\sum_{\lambda_1..._4}\nonumber\\
&\times&F^{\lambda_1,\lambda_2,\lambda_3,\lambda_4}_{\mathbf{k_1,k_1',q}}F^{\lambda_2,\lambda_1,\lambda_4,\lambda_3}_{\mathbf{k_2,k_2',-q}}\nonumber\\
&\times&\Re [\Upsilon_{\lambda_1\lambda_2}(\mathbf{k_1},\mathbf{k_2},\mathbf{q};\epsilon_1^+,\epsilon_2^+)-\Upsilon_{\lambda_1\lambda_2}(\mathbf{k_1},\mathbf{k_2},\mathbf{q};\epsilon_1^+,\epsilon_2^-)]\nonumber\\
&\times& \Re [\Upsilon_{\lambda_3\lambda_4}(\mathbf{k_1'},\mathbf{k_2'},-\mathbf{q};\epsilon_3^+,\epsilon_1-\epsilon_2+\epsilon_3^+)-\Upsilon_{\lambda_3\lambda_4}(\mathbf{k_1'},\mathbf{k_2'},-\mathbf{q};\epsilon_3^-,\epsilon_1-\epsilon_2+\epsilon_3^+)]\nonumber\\
&\times& (1-n_F(\epsilon_1))n_F(\epsilon_2)(1-n_F(\epsilon_3))n_F(\epsilon_1-\epsilon_2+\epsilon_3)
\times(\epsilon_1-\epsilon_2)^2.
\end{eqnarray}

\subsection{Clean case}\label{ap:z2Dirac}
In this case, as $\Upsilon_{\lambda_1, \lambda_2}(\mathbf{k_1},\mathbf{k_2},\mathbf{q};\epsilon_1^\pm,\epsilon_2^+)=\delta(\mathbf{k_1-k_2})G^{R/A}_{\lambda_1}(\mathbf{k_1},\epsilon_1)G^{R}_{\lambda_2}(\mathbf{k_1+q},\epsilon_2)$, the formula for the thermal conductivity is 
\begin{eqnarray}\label{kappaclean}
\kappa& =& 
  \frac{J_\perp^2}{ T^2}\int \frac{d^2k}{(2\pi)^2}\frac{d^2k'}{(2\pi)^2}\frac{d^2q}{(2\pi)^2}
\int\frac{d\epsilon_1}{2\pi}\int\frac{d\epsilon_2}{2\pi}\int\frac{d\epsilon_3}{2\pi}(\epsilon_1-\epsilon_2)^2\sum_{\lambda_1..._4} |F^{\lambda_1,\lambda_2,\lambda_3,\lambda_4}_{\mathbf{k,k',q}}|^2\\
&\times&A_{\lambda_1}(\mathbf{k},\epsilon_1)A_{\lambda_2}(\mathbf{k}+\mathbf{q},\epsilon_2)A_{\lambda_3}(\mathbf{k'},\epsilon_3)A_{\lambda_4}(\mathbf{k'}-\mathbf{q},\epsilon_1-\epsilon_2+\epsilon_3) (1-n_F(\epsilon_1))n_F(\epsilon_2)(1-n_F(\epsilon_3))n_F(\epsilon_1-\epsilon_2+\epsilon_3),\nonumber
\end{eqnarray}
with $A_{\lambda}(\mathbf{k},\epsilon) = -2\Im[G_{\lambda}^R(\mathbf{k},\epsilon)]$ the spinon spectral function, which is $A_{\lambda}(\mathbf{k},\epsilon_1) = 2\pi \delta(\epsilon-\epsilon^\lambda_\mathbf{k})$ in the clean case. This results in the formula derived in the main text, Eq. (\ref{kappaz2clean}).

\subsection{Effects of potential disorder}\label{ap:potdisorder}
For a $Z_2$ QSL with a Fermi surface, we consider the effects of potential disorder. In the self consistent Born approximation (SCBA), which is valid for weak disorder such that $k_F \ell \gg 1$, the dressed Green's function has the form
\begin{eqnarray}
G_\lambda^R(\mathbf{k}, \omega) = G_\lambda^{A*}(\mathbf{k},\omega) = \frac{1}{\omega-\epsilon_{_\lambda\mathbf{k}}+i/2\tau},
\end{eqnarray}
where $\tau = \ell/v_F$ is the disorder-induced lifetime.

In calculating the $4$-point correlation function
\begin{equation}\label{eq:vertexdef}
\Upsilon(\mathbf{k},\mathbf{k}',\mathbf{q};i\nu_n,i\nu_m)
 =\left\langle \psi_l^\dagger(\mathbf{k},i\nu_n)\psi_l(\mathbf{k}+\mathbf{q},i\nu_m)\psi_l^\dagger(\mathbf{k}'+\mathbf{q},i\nu_m)\psi_l(\mathbf{k}',i\nu_n)\right\rangle_\mathrm{dis},
\end{equation}
we define the vertex function $\Gamma_{\mathbf{q}}(i\nu_n,i\nu_m)$ such that
\begin{eqnarray}
\Upsilon(\mathbf{k},\mathbf{k}',\mathbf{q};i\nu_n,i\nu_m) &=& \delta(\mathbf{k-k'})G(\mathbf{k},i\nu_n)G(\mathbf{k+q},i\nu_m) \\
&+&\Gamma_{\mathbf{q}}(i\nu_n,i\nu_m)G(\mathbf{k},i\nu_n)G(\mathbf{k+q},i\nu_m)G(\mathbf{k'},i\nu_n)G(\mathbf{k'+q},i\nu_m)\nonumber
\end{eqnarray}
 In the SCBA, the vertex function is given by the set of ladder diagrams, which are schematically shown in Fig~\ref{fig:vertex}. The sum of all ladder diagrams results in the following self consistent equation for $\Gamma_{\mathbf{q}}(i\nu_n,i\nu_m)$:

\begin{eqnarray}\label{self-consistent}
\Gamma_{\lambda, \lambda}(\mathbf{k}, \mathbf{k+q}; i\nu_n, i\nu_m) &=& \frac{1}{2\pi\nu\tau} +\frac{1}{2\pi\nu\tau}\Gamma_\mathbf{q}(i\nu_n, i\nu_m)\int\frac{d^2k}{(2\pi)^2}G(\mathbf{k},i\nu_n) G(\mathbf{k+q},i\nu_m)\\
&=&\frac{1}{2\pi\nu\tau}\frac{1}{1-\frac{1}{2\pi\nu\tau}\int\frac{d^2k}{(2\pi)^2}G(\mathbf{k},i\nu_n) G(\mathbf{k+q},i\nu_m)}\nonumber
\end{eqnarray}

\begin{figure}
\centering
\includegraphics[width=0.5\textwidth]{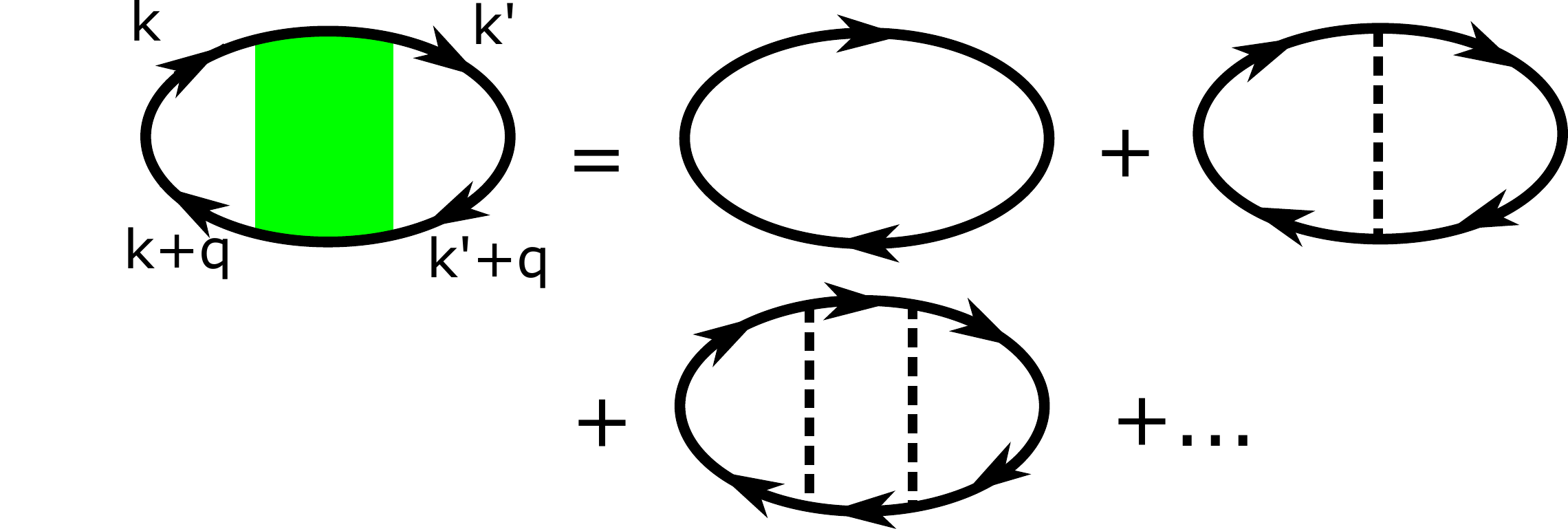}
  \centering
\caption{In the self consistent Born approximation, which is valid when $k_F l_{mfp}\gg 1$, only ladder diagrams without crossed disorder lines contribute to the vertex function. In this figure the full lines are renormalized electron propagators, and dashed lines represent the effects of disorder.}\label{fig:vertex}
    \end{figure}

The important contribution to the thermal conductivity comes from the region of small $\mathbf{q}$ and small frequencies ($\epsilon, v_F\mathbf{q}<T$, with $v_F$ the Fermi velocity). In this region,
\begin{eqnarray}
\Gamma(\mathbf{q}; \epsilon_1^+, \epsilon_2^-) &\approx&\frac{1}{2\pi\nu\tau^2}\frac{1}{-i(\epsilon_2-\epsilon_1)+Dq^2},\nonumber\\
\Gamma(\mathbf{q}; \epsilon_1^+, \epsilon_2^+) &\approx& 0
\end{eqnarray}
with $D =vl_{mfp}/2$ the diffusion constant.

Starting from Eq.~(\ref{kappa}), we perform the Sommerfeld expansion with respect to $\epsilon_1$. The first non-vanishing contribution, in powers of $T$, occurs for the term:
\begin{eqnarray}
\kappa &\sim& 
 {J_\perp^2}\int \frac{d^2k_1}{(2\pi)^2}\frac{d^2k_2}{(2\pi)^2}\frac{d^2k_1'}{(2\pi)^2}\frac{d^2k_2'}{(2\pi)^2}\frac{d^2q}{(2\pi)^2} 
\int\frac{d\epsilon_2}{2\pi}\int\frac{d\epsilon_3}{2\pi}\nonumber\\
&\times&\Re [\Upsilon(\mathbf{k_1},\mathbf{k_2},\mathbf{q};0^+,\epsilon_2^+)-\Upsilon(\mathbf{k_1},\mathbf{k_2},\mathbf{q};0^+,\epsilon_2^-)]\nonumber\\
&\times& \Re [\Upsilon(\mathbf{k_1'},\mathbf{k_2'},-\mathbf{q};\epsilon_3^+,-\epsilon_2+\epsilon_3^+)-\Upsilon(\mathbf{k_1'},\mathbf{k_2'},-\mathbf{q};\epsilon_3^-,-\epsilon_2+\epsilon_3^+)]\nonumber\\
&\times& n_F(\epsilon_2)(1-n_F(\epsilon_3))\partial_\epsilon n_F(-\epsilon_2+\epsilon_3)
\times\epsilon_2^2,\nonumber
\end{eqnarray}
where, at $T\ll E_F$, the last line becomes $n_F(\epsilon_2)(1-n_F(\epsilon_3))\delta(-\epsilon_2+\epsilon_3)
\times\epsilon_2^2 = \delta(-\epsilon_2+\epsilon_3)T\frac{\partial}{\partial\epsilon_2}n_F(\epsilon_2)\epsilon_2^2$. Here we have again set the form factor $\mathcal{F_\mathbf{k,k',q}}=1$  as we are dealing with a single band which crosses the Fermi energy, which is localized on the $A$ sublattice .

After substituting the SCBA result, Eq.~(\ref{self-consistent}), we are left with (neglecting terms with all poles on the same side of the real axis, as these give subleading contributions, and the vertex-less terms, which give a $T^3$ result)

\begin{eqnarray}
\kappa &\sim& 
 {J_\perp^2}T\int \frac{d^2k_1}{(2\pi)^2}\frac{d^2k_2}{(2\pi)^2}\frac{d^2k_1'}{(2\pi)^2}\frac{d^2k_2'}{(2\pi)^2}\frac{d^2q}{(2\pi)^2} 
\int\frac{d\epsilon}{2\pi}\nonumber\\
&\times& \Gamma_\mathbf{q}(0^+,\epsilon^-)G^R(\mathbf{k_1},0)G^A(\mathbf{k_1},\epsilon)G^R(\mathbf{k_2},0)G^A(\mathbf{k_2},\epsilon)\nonumber\\
&\times& \Gamma_\mathbf{q}(\epsilon^+,0^-)G^R(\mathbf{k_1},\epsilon)G^A(\mathbf{k_1},0)G^R(\mathbf{k_2},\epsilon)G^A(\mathbf{k_2},0)\nonumber\\
&\times& \frac{\partial}{\partial\epsilon}n_F(\epsilon)\epsilon^2,\nonumber
\end{eqnarray}

We are interested in the contribution at small $\mathbf{q}$, which has the potential of being singular; we therefore set $\mathbf{q}\rightarrow 0 $ and $\epsilon\rightarrow 0$ in the Green's functions, which results in (Using the relation $G^R(\mathbf{k},\epsilon)G^A(\mathbf{k},\epsilon) = A^2(\mathbf{k},\epsilon) \approx \tau\delta(\epsilon_\mathbf{k}-\epsilon)$, and therefore $\int d^2k/(2\pi)^2 G^R(\mathbf{k},0)G^A(\mathbf{k},0) = \nu\tau$, where $\nu$ is the density of states at the Fermi energy,
\begin{eqnarray}
&&\kappa \sim 
{J_\perp^2}T
\int\frac{d\epsilon}{2\pi}\int\frac{d^2q}{(2\pi)^2}\Re\left[\frac{\nu}{-i\epsilon+Dq^2}\right]\Re\left[\frac{\nu}{i\epsilon+Dq^2}\right]\frac{\partial}{\partial\epsilon}n_F(\epsilon)\epsilon^2\nonumber\\
&& =J^2_\perp\nu^2 T \int\frac{d\epsilon}{2\pi}\int\frac{d^2q}{(2\pi)^2}\left(\frac{Dq^2}{\epsilon^2+D^2q^4}\right)^2\frac{\partial}{\partial\epsilon}n_F(\epsilon)\epsilon^2\nonumber\\
&& = J^2_\perp\frac{\nu^2}{D} T \int\frac{d\epsilon}{2\pi}\frac{\partial}{\partial\epsilon}n_F(\epsilon)|\epsilon|\sim  J^2_\perp\frac{\nu^2}{D} T^2\nonumber
\end{eqnarray}

\subsection{Pair hopping term}\label{ap:pairhopping}
We have ignored the pair hopping term in the paper and in the previous sections. This is because its contribution is similar to that of the spinon-hole hopping term. Performing the Matsubara summation for the pair hopping term results in
\begin{eqnarray}
  &&\frac{J_\perp^2}{ T^2}\int \frac{d^2k_1}{(2\pi)^2}\frac{d^2k_2}{(2\pi)^2}\frac{d^2k_1'}{(2\pi)^2}\frac{d^2k_2'}{(2\pi)^2}\frac{d^2q}{(2\pi)^2} 
\int\frac{d\epsilon_1}{2\pi}\int\frac{d\epsilon_2}{2\pi}\int\frac{d\epsilon_3}{2\pi}\sum_{\lambda_1..._4}\nonumber\\
&&\times\Re [\tilde{\Upsilon}_{\lambda_1\lambda_2}(\mathbf{k_1},\mathbf{k_2},\mathbf{q};-\epsilon_1^+,\epsilon_2^+)-\tilde{\Upsilon}_{\lambda_1\lambda_2}(\mathbf{k_1},\mathbf{k_2},\mathbf{q};-\epsilon_1^+,\epsilon_2^-)]\nonumber\\
&&\times \Re [\tilde{\Upsilon}_{\lambda_3\lambda_4}(\mathbf{k_1'},\mathbf{k_2'},-\mathbf{q};-\epsilon_3^+,\epsilon_1-\epsilon_2+\epsilon_3^+)-\tilde{\Upsilon}_{\lambda_3\lambda_4}(\mathbf{k_1'},\mathbf{k_2'},-\mathbf{q};-\epsilon_3^-,\epsilon_1-\epsilon_2+\epsilon_3^+)]\nonumber\\
&&\times (1-n_F(\epsilon_1))n_F(\epsilon_2)(1-n_F(\epsilon_3))n_F(\epsilon_1-\epsilon_2+\epsilon_3)
\times(\epsilon_1-\epsilon_2)^2,
\end{eqnarray}

where $\tilde{\Upsilon}_{\lambda_1\lambda_2}(\mathbf{k_1},\mathbf{k_2},\mathbf{q};i\nu_n,i\nu_m) = \langle a_{\lambda_1}(-\mathbf{k_1},-i\nu_n)a_{\lambda_2}(\mathbf{k_1+q},i\nu_m)a^\dagger_{\lambda_2}(\mathbf{k_2+q},i\nu_m)a^\dagger_{\lambda_1}(-\mathbf{k_2},i\nu_n)\rangle$.

An analysis similar to that following Eq.~(\ref{eq:vertexdef}) shows that
\begin{equation}
\begin{split}
\tilde{\Upsilon}_{\lambda_1\lambda_2}(\mathbf{k_1},\mathbf{k_2},\mathbf{q};-\epsilon_1^+,\epsilon_2^-) &= \delta(\mathbf{k_1}-\mathbf{k_2})G_{\lambda_1}^R(\mathbf{-k_1},\epsilon_1)G^A_{\lambda_2}(\mathbf{k_1+q},\epsilon_2)\\
&+\Gamma_\mathbf{q}(\epsilon_1^+,\epsilon_2^-)G_{\lambda_1}^R(\mathbf{-k_1},\epsilon_1)G_{\lambda_2}^A(\mathbf{k_1+q},\epsilon_2)
G_{\lambda_1}^R(\mathbf{-k_2},\epsilon_1)G_{\lambda_2}^A(\mathbf{k_2+q},\epsilon_2)\\
\tilde{\Upsilon}_{\lambda_1\lambda_2}(\mathbf{k_1},\mathbf{k_2},\mathbf{q};-\epsilon_1^+,\epsilon_2^+) &= \delta(\mathbf{k_1}-\mathbf{k_2})G_{\lambda_1}^R(\mathbf{-k_1},\epsilon_1)G^R_{\lambda_2}(\mathbf{k_1+q},\epsilon_2),
\end{split}
\end{equation}

with $\Gamma_\mathbf{q}(\epsilon_1^+,\epsilon_2^-) = \frac{1}{2\pi\nu\tau^2}\frac{1}{-i(\epsilon_2-\epsilon_1)+Dq^2}$ as before; this term therefore contributes the same as the spinon-hole hopping term.


\section{$U(1)$ Quantum Spin Liquid}\label{ap:U1}
\subsection{Clean spinon thermal conductivity}
In this case, the interlayer coupling is given by Eq.~(\ref{couplings})
\begin{eqnarray}
H_{\perp,sp} 
 &=&J^{sp}_{\perp}\int \frac{d^2k d^2k' d^2q}{(2\pi)^6}\psi_l^*(\mathbf{k})\psi_l(\mathbf{k+q})\psi_{l'}^*(\mathbf{k'})\psi_{l'}(\mathbf{k'-q});
\end{eqnarray}
Plugging this into the formula for the thermal current operator Eq.~(\ref{JQ}), and using the Kubo formula just as in Eq.~( \ref{kuboz2}), results in
\begin{eqnarray}
\kappa_{sp} = &&\frac{J_{\perp}^{sp 2}}{T^2}\int \frac{d^2k d^2k' d^2q}{(2\pi)^6}\int \frac{d\epsilon_1d\epsilon_2d\epsilon_3}{(2\pi)^3}A(\mathbf{k}, \epsilon_1)A(\mathbf{k+q}, \epsilon_2)A(\mathbf{k'}, \epsilon_3)A(\mathbf{k'-q},\epsilon_1-\epsilon_2+\epsilon_3)
\times\nonumber\\
&&(\epsilon_1-\epsilon_2)^2n_F(-\epsilon_1)n_F(\epsilon_2)n_F(-\epsilon_3)n_F(\epsilon_1-\epsilon_2+\epsilon_3),
\end{eqnarray}
with $A(\mathbf{k}, \epsilon)$ the spinon spectral function 
\begin{eqnarray}
A(\mathbf{k}, \epsilon) = \frac{2c\epsilon^{2/3}}{(\epsilon_\mathbf{k}-\mu)^2+c^2\epsilon^{4/3}},\\
A(\mathbf{k}, \epsilon=0) = 2\pi\delta(\epsilon_\mathbf{k}-\mu)\nonumber,
\end{eqnarray}
and $c = (k_F/m)\chi_D^{-2/3}k_0^{-1/3}$. \cite{Motrunich, Senthil, LeeReview, Ioffe, Nagaosa, Lee2, Polchinski}
We consider the contribution of $q\ll k_F$, expanding $\epsilon_\mathbf{k+q}\approx v_F|\mathbf{k+q}|$; we then apply the Sommerfeld expansion according to $\epsilon_1$, and the largest contribution at low $T$ comes from the term where the derivative is applied to $n_F(\epsilon_1-\epsilon_2+\epsilon_3)$
\begin{eqnarray}
\kappa_{sp} \approx J_{\perp}^{sp 2}\int \frac{d^2k d^2k' d^2q}{(2\pi)^6}\int \frac{d\epsilon_2d\epsilon_3}{(2\pi)^2}A(\mathbf{k}, 0)A(\mathbf{k+q}, \epsilon_2)A(\mathbf{k'}, \epsilon_3)A(\mathbf{k'-q},-\epsilon_2+\epsilon_3)
\epsilon_2^2n_F(\epsilon_2)n_F(-\epsilon_3)\partial_{\epsilon}n_F(-\epsilon_2+\epsilon_3),\nonumber\\
\end{eqnarray}
which at low temperatures becomes
\begin{eqnarray}
\kappa_{sp} \approx J_{\perp}^{sp 2}T\int \frac{d^2k d^2k' d^2q}{(2\pi)^6}\int \frac{d\epsilon_2}{2\pi}A(\mathbf{k}, 0)A(\mathbf{k+q}, \epsilon_2)A(\mathbf{k'}, \epsilon_2)A(\mathbf{k'-q},0)
\frac{\partial n_F(\epsilon_2)}{\partial\epsilon_2}\epsilon_2^2.\nonumber\\
\end{eqnarray}
Using 
\begin{eqnarray}
\int \frac{d^2k}{(2\pi)^2}A(\mathbf{k},0)A(\mathbf{k+q},\epsilon) = \nu\int d\theta \frac{c\epsilon^{2/3}}{v_F^2q^2\cos^2\theta+c^2\epsilon^{4/3}} = \frac{\nu}{\sqrt{v_F^2q^2+c^2\epsilon^{4/3}}}
\end{eqnarray}
we find that
\begin{eqnarray}
\kappa_{sp} \approx J_{\perp}^{sp 2}\nu^2 T\int \frac{d\epsilon}{2\pi}\frac{\partial n_F(\epsilon)}{\partial\epsilon}\int \frac{d^2q}{(2\pi)^2}\frac{\epsilon^{2}}{v_F^2q^2+c^2\epsilon^{4/3}} \approx \frac{J_{\perp}^{sp 2}\nu^2}{v_F^2}T^3\log\left(\frac{T}{(W/c)^{3/2}}\right)
\end{eqnarray}
where $W$ is a UV cut-off. 

\subsection{Clean gauge photon thermal conductivity}
In this case, the coupling of the interlayer gauge-fields is given by Eq.~(\ref{U1couplings})
\begin{eqnarray}
H_{\perp,ph}  &=&  J^{ph}_{\perp}\int\frac{d^2k}{(2\pi)^2}k^2a^T_l(k)a^T_{l'}(k'),
\end{eqnarray}

and therefore the thermal conductivity is
\begin{eqnarray}
\frac{(J^{ph}_{\perp})^2}{T}\int \frac{d^2k}{(2\pi)^2} k^4 \int_0^\infty d\epsilon A_{ph}^2(\mathbf{k},\epsilon)\epsilon^2\partial_\epsilon n_B(\epsilon),
\end{eqnarray}
where $A_{ph}(\mathbf{k},\epsilon) = \gamma\frac{|\omega|k}{\chi^2 k^6+\gamma^2\omega^2}$ is the photon spectral function. This results in 
\begin{eqnarray}
\kappa &=& \frac{\gamma^2(J^{ph}_{\perp})^2}{T}\int \frac{d^2k}{(2\pi)^2} k^6 \int_0^\infty d\epsilon \frac{\epsilon^4}{\left(\chi^2 k^6 + \gamma^2\epsilon^2\right)^2}\partial_\epsilon n_B(\epsilon)\\
&\sim&  \frac{\gamma^2(J^{ph}_{\perp})^2}{T} \left(\frac{\gamma}{\chi}\right)^{-4/3}\int_0^\infty d\epsilon \epsilon^{8/3}\partial_\epsilon n_B(\epsilon)\sim (J^{ph}_{\perp})^2\gamma^{2/3}\chi^{4/3}T^{5/3}.\nonumber
\end{eqnarray}

In our calculation of the interlayer thermal conductivity, we have neglected processes which transfer a larger number of gauge-invariant excitations between the layers (for example, two spinons and a photon). This is because their contribution to $\kappa_c$ has a higher power of T and is therefore negligible in the limit of low temperature. 

\end{document}